# Localized Edge States in Stacked Al/Ni Multilayers: Possible Evidence of Chiral Hinge Modes

M. Belogolovskii[1]* and I. P. Nevirkovets[2]

[1] Centre for Nanotechnology and Advanced Materials, Faculty of Mathematics, Physics and Informatics, Comenius University Bratislava, Mlynská dolina, Bratislava 84248, Slovak Republic
[2] Department of Physics and Astronomy, Northwestern University, 2145 Sheridan Road, Evanston, Illinois 60208, USA

## Abstract

Here, we report experimental evidence suggesting the emergence of robust, possibly chiral, edge states in artificially engineered multilayers composed of alternating nanometer-thick layers of nonmagnetic aluminum (Al) and ferromagnetic nickel (Ni). Using phase-sensitive Josephson interferometry, we observed distinct SQUID-like oscillations (instead of the conventional Fraunhofer patterns) in the maximum supercurrent versus in-plane probing magnetic field patterns, which can be associated with one-dimensional current-carrying modes localized at the sample boundaries. These results were obtained for multilayers consisting of up to ten Al/Ni bilayers sandwiched between superconducting Nb electrodes to form Josephson junctions. The spatially confined flow of supercurrent suggests the possible presence of chiral Andreev edge states reminiscent of those found in higher-order topological insulators, despite the absence of strong spin-orbit coupling or intrinsic topological band structure. The discovery of edge-localized charge transport in structures made of materials without intrinsic topological order challenges the prevailing understanding of topological phenomena and highlights the possibility of developing topological metamaterials as a tunable platform for exploring nontrivial edge physics.

----------------

*Contact author: belogolovskii@ukr.net

# I. INTRODUCTION

In Josephson junctions (JJ) with a non-superconducting (N) weak link sandwiched between two *s*-wave superconducting (S) electrodes, the transport of Cooper pairs through the N metallic barrier is realized via specific (Andreev) backscattering processes at the N/S boundaries in which an incident electron is retroreflected as a hole [1]. In conventional Josephson SNS devices, incoming electrons typically move more or less uniformly within the N-layer, and holes come back along their trajectories [2-4]. A way to test this scenario and prove the Josephson character of charge transport through such weak links is to measure the maximum supercurrent $I_c$ as a function of a *probing* magnetic field $H$ applied parallel to the layers and hence perpendicular to the current direction, making the associated difference in the gauge-invariant quantum phases a spatially dependent quantity. For traditional SNS and SIS (I is a nanometer-thin insulating layer) samples, we obtain the well-known $I_c(\Phi)$ dependence, dubbed the Fraunhofer pattern, with a central lobe of width $2\Phi_0$ and decaying side lobes with nodes located at the $\Phi_0$ distance from each other [2]. Here $\Phi_0 = hc/2e$ is the flux quantum, $h$, $c$, and $e$ are the Planck constant, the speed of light, and the electron charge, respectively; $\Phi$ is the flux threading the weak link, the product of the field $H$ and the sensing area $s$, defined below.

When the current is distributed unevenly across the weak link, for example, the transfer of Cooper pairs through a Josephson junction is possible only through the edge states, the $I_c(\Phi)$ characteristic is expected to be fundamentally different from the Fraunhofer one. In this case, similar to the Aharonov-Bohm effect, the measured maximum supercurrent $I_c$ in response to the external magnetic flux $\Phi$ oscillates with the probing magnetic field exhibiting a SQUID-like quantum interference pattern with a certain periodicity (compare Figs. 1b and 1a in Ref. [5]). For helical edge modes, when electrons and retroreflected holes are flowing along the same edge, the period of the $I_c(\Phi)$ pattern equals to $\Phi_0$. However, Andreev scattering processes become more complicated for one-way (*chiral*) edge currents. In such weak links, electron-to-hole (and inverse) backscattering described above is not allowed, and the transfer of Cooper pairs through the JJ is carried out by electron and hole states on the opposite edges of the sample linked via crossed Andreev reflections [6]. As a result, we get an $I_c(\Phi)$ curve with two distinct features: (i) a periodic dependence determined by a normal flux quantum $hc/e$ instead of $\Phi_0 = hc/2e$, usual for superconducting JJs [6] and (ii) a pronounced shift of the SQUID-like $I_c(\Phi)$ characteristic upward along the maximum supercurrent $I_c$ axis (Fig. 1c in Ref. 5). The latter result is an evident consequence of the spatial asymmetry of the currents flowing through the weak link and localized along its edges.

It is known that helical or chiral edge current-carrying states can be realized by using topological materials as weak links [5]. In this case, the Josephson effect can serve as an extremely useful probe for edge superconductivity in the hybrid topological phases of matter, which are of considerable interest for both fundamental research and advanced technologies [7]. The distinctive property of topological solid-state materials is the fundamental difference between their bulk and surface electronic characteristics. According to the bulk-boundary correspondence, a conventional topological insulator (TI) is insulating in the bulk, while its surface hosts conducting states, meaning that electrons can move only along the superficial shell [8]. The formation of the gapless surface state together with a bulk band gap is a result of the strong spin-orbit coupling responsible for protecting the surface states from backscattering, thus making them robust to disorder and backscattering [9]. The unique electronic structure of this class of materials, known as first-order topological insulators [10], has led to numerous potential applications in quantum computing, spintronics, and other prospective fields [11].

Recent theoretical and experimental developments have revealed the existence of a more complex class of materials, dubbed higher-order topological insulators (HOTIs) with the co-dimension $n > 1$, for which a $d$-dimensional $n$th-order topological phase supports $(d-n)$-dimensional boundary states [10,12]. For the three-dimensional ($d = 3$) case, this means that, unlike first-order TIs, where conductivity occurs over the entire surface, second-order TIs (SOTIs) contain one-dimensional (1D) conducting modes localized along the hinges of the structure [10]. The concept of SOTIs offering exciting possibilities for dissipationless transport and novel quantum phenomena has opened new avenues in topology-related research. As a result, the search for materials exhibiting hinge states has become a rapidly growing field in condensed matter physics, with experimental confirmations in bulk bismuth, bismuth bromide, and transition-metal dichalcogenides, see the references in the overview [12]. Depending on the presence or absence of time-reversal symmetry, there can be two types of hinge conductive modes: Kramers pairs of helical hinge states protected by an additional crystalline symmetry (mirror or rotational symmetry) and chiral hinge states arising from the breaking of time-reversal symmetry and protected by the product of time-reversal and rotational symmetry [12].

Therefore, 3D *chiral* SOTI should contain gapped bulk and side surface states and at the same time 1D topologically protected in-gap hinge chiral modes [12,13]. In addition to searching for new crystalline materials with intrinsic SOTI phases, another route for the development of such topological objects consists of layer stacking with interlayer coupling tuning. It has now become a generally accepted method, demonstrating its ability to create new topological phases in solids [14] and photonic systems [15-18]. Theoretical analysis of the effect of quenched disorder on the higher-order topology showed that, similar to first-order topological phases, the 3D chiral SOTIs remain robust to a moderate decrease in the elastic mean free time of the quasiparticles [13]. Even more, it was found that SOTIs with either chiral or helical in-gap hinge states could exist in specific 3D amorphous systems characterized by nontrivial topological invariants [19].

As stated above, the insertion of such materials inside the Josephson junction, followed by the application of an in-plane probing magnetic field $H$ and subsequent analysis of the $I_c(H)$ patterns, makes it possible to unambiguously determine the nature and dimensionality of charge transport through the objects under study. Despite a large number of materials used as weak links [20], multilayer structures of alternating nanometer-thick films of normal and ferromagnetic (F) metals, fully corresponding to the above-described stacking technology for constructing SOTI photonic phases [14-18], have not been studied in detail. With this in mind, we designed a series of Josephson junctions based on N/F multilayers to explore how the supercurrent is distributed and whether the $I_c(H)$ curve follows the expected Fraunhofer behavior for conventional metallic systems.

The electronic versions of stacked heterostructures were formed by periodic layering N/F pairs of non-superconducting paramagnetic N = Al and ferromagnetic F = Ni or permalloy metals, Fig. 1a. The Josephson interferometric methodology was used to study the transport of Cooper pairs through an $(NF)_n$N multilayer composed of up to $n$ = 10 metallic N/F bilayers. To our surprise, instead of the conventional Fraunhofer pattern, we observed SQUID-like $I_c$ oscillations as a function of the in-plane magnetic field [21]. This finding was explained by hypothesizing the presence of edge currents in the hybrid weak link, while the question of their directionality (one- or two-way transport) has not been discussed. Keeping details in the Appendices, we present below such an analysis for JJs with an $(NF)_{10}$NI$(NF)_{10}$N hybrid barrier confirming the assumption on the dominant edge supercurrent transport with chirality features, Fig. 1b. Furthermore, we extended our research to more complicated multilayered weak links in which the oxide I interlayer was replaced by a conventional Josephson S′IS′ trilayer (S′ is an ultrathin Nb film). The latter experiments were aimed at clarifying the details of the transformation of the edge currents upon their contact with a superconducting plane. This approach allowed us to distinguish the specific Andreev conversion in chiral modes from the conventional Andreev retroreflection in helical states, which is a non-trivial task for topological systems [22].

A comprehensive interpretation of the findings presented in the work convincingly points to the emergence of chiral Andreev edge (possibly hinge) modes in the hybrid $(NF)_n$ multilayers with a relatively large number of bilayers $n \cong 10$. The observed characteristics strongly resemble those seen in Josephson junctions with specific materials in quantum Hall and anomalous quantum Hall regimes in spite of the fact that we are dealing with *ordinary* and well-studied metallic films. Moreover, unlike previous experiments, using Josephson interferometry of *intrinsically topological* materials to reveal quantum transport in the gapless topological edge or hinge states [5], our samples are not planar junctions but rather conventional Josephson multilayers. Other details of the study, including additional measurements and theoretical considerations, are provided in the Appendices.

## II. EXPERIMENT

To keep the critical current of the SNS-type JJs with a normal interlayer at a sufficiently low level, their lateral dimensions should be quite small. There is no such restriction for low-transparency SIS devices with oxide barriers. Because of this, we fabricated and studied 10 μm × 10 μm and 5 μm × 5 μm $(NF)_{10}$-based JJs with an $AlO_x$ interlayer of three types of hybrid multilayers: $S(NF)_{10}NI(NF)_{10}NS$, $S(NF)_{10}NS'IS'(NF)_{10}NS$, and $S(NF)_{10}NS'IS$ with S = S′ = Nb. N = Al, F = Ni, and I = Al/$AlO_x$. This choice is not fundamental, since both types of samples, with and without the oxide barrier, behave qualitatively in the same way, see our publication [21]. A brief overview of our previous results (published and unpublished) on JJs with different numbers of N/F bilayers is given in Appendix D. Specifically, we present the evolution of the $I_c(H)$ curves as the number $n$ of bilayers forming the JJ's weak link increases from $n = 0$ (the well-known Fraunhofer pattern observed in SIS devices, Fig. S4 in the Supplementary Information) to $n = 10$ (the SQUID-like characteristic with a slowly decaying envelope, Fig. 1c below and Fig. A4 in Appendix D).

The Josephson junctions were fabricated from corresponding hybrid multilayers deposited *in situ* onto oxidized Si substrates using DC magnetron sputtering of the corresponding targets at room temperature. The insulating Al-oxide interlayer was formed on the Al film by thermal oxidation. The cross-type Josephson devices were patterned using optical lithography, reactive ion etching, Ar ion milling, and anodization, followed by the deposition of additional $SiO_2$ insulation. Let us note that all samples with N/F-bilayer numbers from $n = 1$ to $n = 10$ were prepared using the same fabrication technique, which, in particular, involved anodizing the junction edges for insulation. The junctions with $n = 1, 2$ exhibited the standard Fraunhofer-like $I_c(H)$ interference patterns (though sometimes shifted in $H$, see Appendix D) with $I_c = 0$ minima, implying that edge current-carrying channels are caused by the collective behavior of the constituent elements of the multilayers with $n > 2$.

Thicknesses of the bottom and top Nb layers were 120 and 80 nm, respectively, while the S′ niobium films forming inner Josephson junctions within the complex multilayers were as thin as 12 nm. The ultrathin Al and Ni films constituting stacking $(Al/Ni)_n$ periods were 3.1 and 1.3 nm thick, respectively. After depositing a thick (~ 400 - 500 nm) Nb wiring layer followed by lift-off, the Josephson devices were completed for four-probe measurements at the temperature of 4.2 K. Schematic views of the three types of Josephson junctions studied in the work are shown in Figs. 1a, 2a, and 3a.

In Fig. A1, we provide a cross-sectional TEM (transmission electron microscopy) image of a representative $Nb[Al/Ni]_{10}Al/AlO_x[Al/Ni]_{10}Al/Nb$ device showing well-resolved Ni and Al layers within the periodic $S(N/F)_{10}NI(N/F)_{10}NS$ stack. A higher-resolution TEM analysis of our three-terminal devices comprising Py and Al layers performed at Sandia National Laboratories indicated the presence of some interdiffusion between Py and Al that can weaken the F-layer magnetic properties in the multilayered samples.

To characterize the magnetic state of our $(Al/Ni)_n$ multilayers, a large-area $Nb/(Al/Ni)_{70}$ sample of 5.5×11.4 $mm^2$ was fabricated on a $Si/SiO_2$ substrate with the same thicknesses of Al and Ni films as above. Next, the magnetic moment $M$ as a function of $H$ applied parallel to the films was measured at 10 K, just above the critical temperature $T_c \approx 9$ K of niobium. The analysis presented in the paper [21] showed that we are dealing with an *artificial superparamagnet* consisting of alternating nm-thick layers of ferromagnetic and non-magnetic metals. In this case, one expects that the magnetic moments of the superparamagnetic clusters are switching between the two directions of the easy axis, creating magnetic domains *along* the junction interface. Such expectation is supported by numerical simulations [23] for a thin ferromagnetic film with strong uniaxial anisotropy and perpendicular to its plane's easy axis. Starting from the full three-dimensional micromagnetic model, the authors of Ref. [23] found a transition from a single-domain to a multi-domain ground state occurring as the film thickness tends to zero while the lateral dimensions tend to infinity. If so, we can expect the emergence of the time-reversal symmetry breaking, which in turn leads to an increase in the skew (asymmetric) scattering potential previously observed in Ref. [24] for single polycrystalline Ni films with thicknesses varying from 7 nm to 15 nm.

Two length scales, the coherence length $\xi$ and the magnetic-field penetration depth $\lambda$, are the basic characteristics of superconducting materials. A very brief literature analysis of $\lambda_{Nb}$ quantities presented in Appendix B indicates that its value has to be found for each specific technology of the Josephson junction fabrication. Therefore, we determined $\lambda_{Nb}$ using the maximum supercurrent $I_c$ versus the in-plane magnetic field $H$ dependence for a conventional SIS junction (S = Nb), where $\lambda_{Nb}$ and the $I_c(H = 0)$ amplitude were two fitting parameters. We have found that the London penetration depth in our Nb films $\lambda_{Nb}$ at 4.2 K is equal to 82 nm. Our superconducting Nb electrodes are in the dirty limit when the electron mean free path $l$, controlled by scattering on impurities, is considerably smaller than the coherence length $\xi_0$ in pure and bulk Nb at 0 K. From resistance-temperature measurements, it is possible to determine the value of $l$ knowing that its product with the bulk resistivity $\rho$ is equal to $\rho l = 3.75\cdot10^{-6}$ μΩ·cm$^2$ [25]. For our Nb films, $\rho = 5$ μΩ·cm [26], hence, $l \approx 7.5$ nm at low temperatures. Using the formula for the effective coherence length $\xi(T) = 0.85(\xi_0 l)^{1/2}/(1-T/T_c)^{1/2}$, we find that $\xi(T = 4.2\text{ K}) \approx 20$ nm.

Current-voltage $I(V)$ characteristics and the dependences of the maximum Josephson current $I_c$ on the external magnetic field $H$ applied parallel to the metallic layers [the $I_c(H)$ patterns] were measured in a liquid He bath using the four-probe configuration. The measurements were carried out in a shielded room using battery-powered electronics to reduce the noise. It is well known [27] that the $I_c(H)$ dependence and the spatial distribution of the $I_c$ density, $J_s(y)$, are related via the Fourier transform, which allows the extraction of the $J_s(y)$ spatial behavior in the weak link of a Josephson junction from the measured $I_c(H)$ pattern. Details of the procedure are outlined in Appendix C. Reconstructed supercurrent density profiles are shown in Figs. 1d, 2d, and 3d for the 10 μm × 10 μm samples.

# III. RESULTS

## A. Josephson junctions studied

In this work, we consider three types of JJs with lateral dimensions of 10 μm × 10 μm: Nb(120)[Al(3.1)Ni(1.2)]$_{10}$Al/AlO$_x$(3.1)[Al(3.1)Ni(1.2)]$_{10}$Al(3.1)Nb(75) [S(NF)$_{10}$NI(NF)$_{10}$NS], Nb(120)[Al(3.1)Ni(1.3)]$_{10}$Al(9.3)Nb(12)Al/AlO$_x$(9.3)Al(3.1)Nb(12)[Al(3.1)Ni(1.3)]$_{10}$Al(3.1)Nb(80) [S(NF)$_{10}$NS′IS′(NF)$_{10}$NS], Nb(120)[Al(3.1)Ni(1.3)]$_{10}$Al(9.3)Nb(12)Al/AlO$_x$(9.3)Al(3.1)Nb(80) [S(NF)$_{10}$NS′IS]. The numbers in parentheses indicate the nominal thicknesses of respective layers in nm; as for Al/AlO$_x$, this is the thickness of the Al overlayer before thermal oxidation. The thickness of the initially deposited top Nb electrode did not take into account the presence of an additional niobium wiring layer, the inclusion of which increased the overall thickness of the top electrode to approximately 500 nm. Figs. 1a, 2a, and 3a schematically show the structure of the studied devices (not to scale) with their biasing, while Figs. 1b, 2b, and 3b illustrate the physical processes at interfaces using semi-classical skipping orbit images.

Below we consider the current-voltage ($I$-$V$) characteristics and the dependence of the maximum Josephson current $I_c$ on the external magnetic field $H$ applied along the $x$-axis in the $xy$-plane parallel to the layers; see the inset in Fig. 1a. Dynes and Fulton [27] showed that the $I_c(H)$ dependence and the spatial distribution of the supercurrent density, $J_s(y)$, are related via the Fourier transform. This allows one to reconstruct the spatial behavior of $J_s(y)$ in a weak link from the measured $I_c(H)$ pattern. Details of the reconstruction procedure are outlined in Appendix C.

As is mentioned in the Introduction, an important characteristic of a Josephson junction is the sensing area $s$, a product of its width $w$ and the effective magnetic thickness $t$ that describes the magnetic field penetration into a weak link and adjacent regions of the S electrodes. With the barrier magnetic permeability $\mu$, its thickness $d_B$, and electrode thicknesses $d_{S1}$ and $d_{S2}$ comparable to the London penetration depths $\lambda_1$ and $\lambda_2$, the formula for $t$ is as follows [28,29]:

$$t = \mu d_B + \lambda_1 \tanh(d_{S1}/2\lambda_1) + \lambda_2 \tanh(d_{S2}/2\lambda_2). \tag{1}$$

Here $\lambda_1 = \lambda_2 = \lambda_{Nb} = 82$ nm.

### B. S(NF)$_{10}$NI(NF)$_{10}$NS junctions

Let us start with the analysis of experimental data for a representative device Nb(120)[Al(3.1)Ni(1.2)]$_{10}$Al/AlO$_x$(3.1)[Al(3.1)Ni(1.2)]$_{10}$Al(3.1)Nb(470) with an Al/AlO$_x$ insulating barrier that separates two (NF)$_{10}$N multilayers. Its schematic view is shown in Fig. 1a where for simplicity we display the S(NF)$_n$NI(NF)$_n$NS device with $n$ = 5. Typical $I_c(H)$ dependence of an S(NF)$_{10}$NI(NF)$_{10}$NS JJ, shown in Fig. 1c, demonstrates a well-defined supercurrent and upwardly displaced SQUID-like oscillations with a smooth and slowly decreasing envelope function, i.e., the features which, as noted above, are characteristics of devices formed by topological materials with chiral edge currents sandwiched between two superconductors,

The period of field oscillations following from Fig. 1c is $\Delta H \approx 7.8$ Oe. To compare it with the related value for the magnetic flux period $\Delta\Phi$, we need to estimate the effective magnetic thickness $t$ from Eq. (1). Setting $\mu \approx 5$ obtained previously for (Al/Ni)$_n$ multilayers [21], we find $t$ = 560 nm, the sensing area s = $5.6\cdot10^{-8}$ cm$^2$ and, finally, the experimental value $\Delta\Phi = \Delta H{\cdot}s = 4.4\cdot10^{-7}$ G·cm$^2$ which is approximately twice the magnetic flux quantum $\Phi_0 = 2.068\cdot10^{-7}$ G·cm$^2$. Leaving the discussion of the possible spread of the magnetic permeability value in each specific experiment with an N/F multilayer until the next section (see also Appendix D), we propose a possible fundamental reason for the period doubling using a semi-classical interpretation of alternating electron-to-hole transformations shown schematically in Fig. 1b. When an N-layer with chiral edge currents approaches the superconductor, an electron incident on the N/S interface from one edge of the N-barrier emerges as a hole on the other side after multiple Andreev retroreflections along the S surface. If this occurs inside a JJ, then, as was already stated in the Introduction, the emerging *chiral Andreev edge states* hybridizing electron and hole amplitudes lead to a *doubling* of the period of $I_c(H)$ oscillations [6].

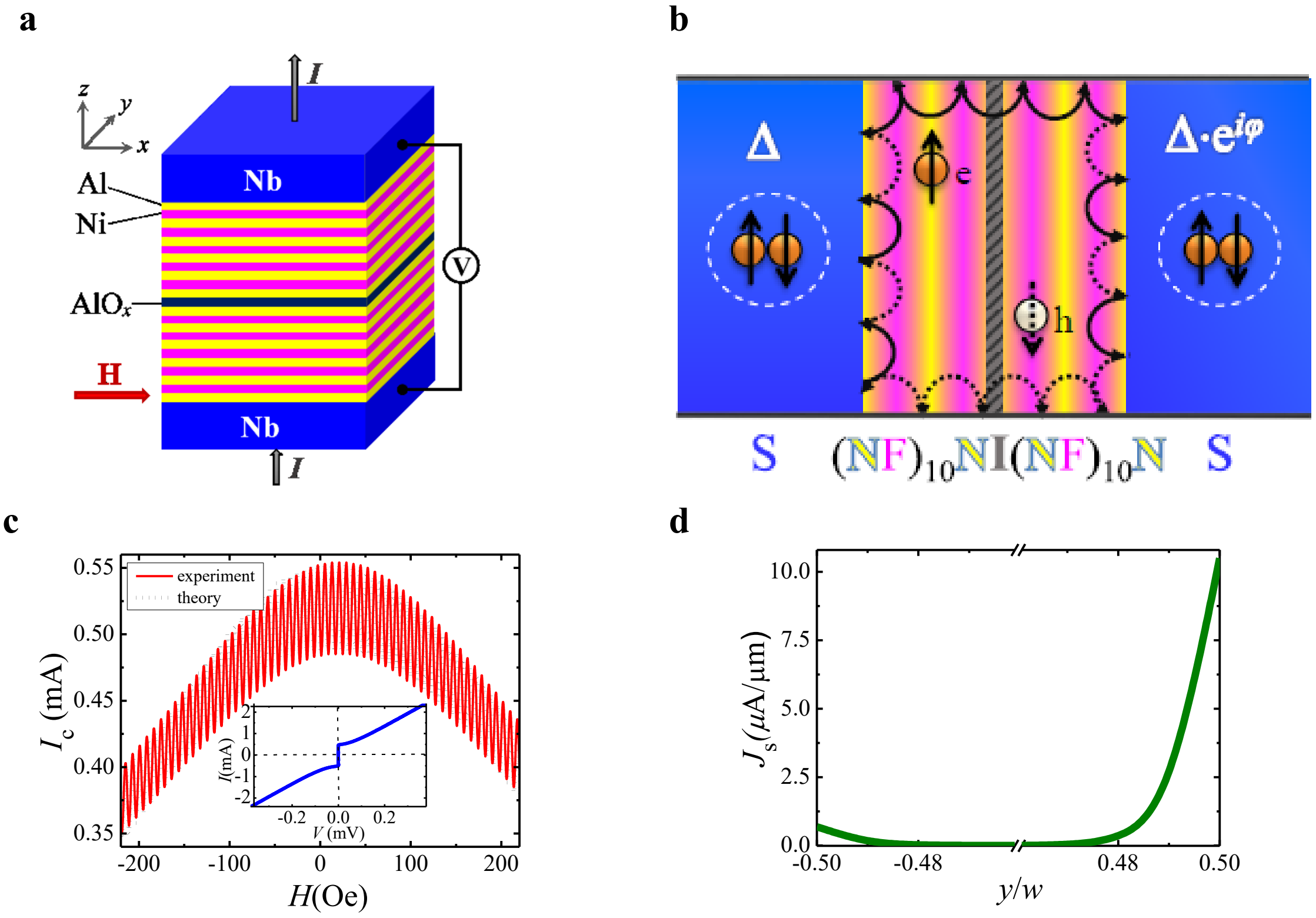

**Fig. 1.** S(NF)$_{10}$NI(NF)$_{10}$NS multilayer. **a**, Schematic view of a 10 μm × 10 μm S(NF)$_n$NI(NF)$_n$NS Josephson junction (not to scale) and its biasing. For simplicity, we show a heterostructure with $n$ = 5, but not 10, as it was. **b**, Semi-classical skipping orbit image of charge trajectories in the *yz*-plane, which are confined to the side surfaces of the multilayered barrier and the interfaces with superconducting electrodes. **c**, Main panel: a measured pattern (red solid line) and the theoretical (black dots) $I_c(H)$ dependence calculated using the supercurrent density profile displayed in Fig. 1d. The inset shows the measured *I-V* curve. **d**, Extracted spatial behavior of the supercurrent density profile $J_s(y)$ integrated over the *x* direction. The measurement temperature was 4.2 K.

The assumption about chiral currents inside the (NF)$_n$NI(NF)$_n$N multilayer can be supported by extracting the spatial dependence of the supercurrent density $J_s(y)$ using the method described in Appendix C. Fig. 1d indeed confirms the predominantly chiral nature of the currents in the barrier. At the same time, it should be noted that the degree of dominance of a certain direction in the charge motion inside the multilayer barrier over the reverse one varies significantly from sample to sample even within the same batch of junctions, which nominally should be identical, see Fig. A7 in Appendix D, which shows the $I_c(H)$ patterns for six 5 μm × 5 μm Nb(120)[Al(3.1)Ni(1.2)]$_{10}$Al(3.1)Nb(470) junctions. The significant spread in the amplitude values of the $I_c(H)$ oscillations in Fig. A7 with their practically unchanged period can be understood by assuming the presence of 1D intrinsic edge modes with oppositely directed currents, the relationship between which is determined by random magnetic factors in each channel with the final result depending on their balance.

### C. S(NF)$_{10}$NS′IS′(NF)$_{10}$NS junctions

As follows from Fig. 1b, the charge motion along the S surface is accompanied by multiple Andreev reflections, which leads to the penetration of Cooper pairs into the superconductor and a much more even distribution of the supercurrent density in the S electrode compared to that within the multilayer, Fig. 1d. One of the ways to test this claim is to replace the insulating I interlayer in the S(NF)$_{10}$NI(NF)$_{10}$NS multilayer with a standard S′IS′ Josephson trilayer, Fig. 2a. We expected that the S′IS′ junction inside the multilayer weak link would serve as a "bottleneck" controlling the overcurrent throughout the entire device. Therefore, measuring the $I_c(H)$ dependence for the entire sample would allow us to determine whether the supercurrent flows through the S′IS′ trilayer along its edges, as is the case for helical modes, or spreads over the entire area of the inner S′IS′ junction, as is the case for chiral states, Fig. 2b. In the latter case, the conventional Fraunhofer pattern should be observed instead of the SQUID-like oscillations like those in Fig. 1c. Such a test was implemented on the Nb(120)[Al(3.1)Ni(1.3)]$_{10}$Al(3.1)Nb(12)Al/AlO$_x$(9.3)Al(3.1)Nb(12)[Al(3.1)Ni(1.3)]$_{10}$Al(3.1)Nb(480) heterostructures. The main panel of Fig. 2c shows the measured $I_c(H)$ dependence for a S(NF)$_{10}$NS′IS′(NF)$_{10}$NS device, while the inset demonstrates its *I-V* curve. The pronounced hysteresis in the *I–V* characteristics, which is absent in the S(NF)$_{10}$N′I(NF)$_{10}$NS samples (cf. the inset in Fig. 1c), confirms the dominance of charge transfer through the S′IS′ junction.

The shape of the measured $I_c(H)$ dependence [see Fig. 2c] is indeed close to that of the Fraunhofer pattern, but with an unusually large period $\Delta H \approx 102$ Oe due to the very small thickness (12 nm) of the S′ electrodes. As is known, the magnetostatic interaction between a comparatively large ferromagnetic layer with an in-plane magnetization and an adjacent superconducting film is practically absent, hence, we can assume that the intrinsic magnetic field is concentrated predominantly inside the (NF)$_{10}$N multilayers and take $\mu = 1$ for the I barrier. As a result, we get the sensing area of the S′IS′ junction $s = 0.24$ μm$^2$. The obtained period $\Delta\Phi = 2.4\cdot10^{-7}$ G·cm$^2$ is in good agreement with the magnetic flux quantum $\Phi_0 = 2.068\cdot10^{-7}$ G·cm$^2$ as expected for the conventional SIS Josephson junctions.

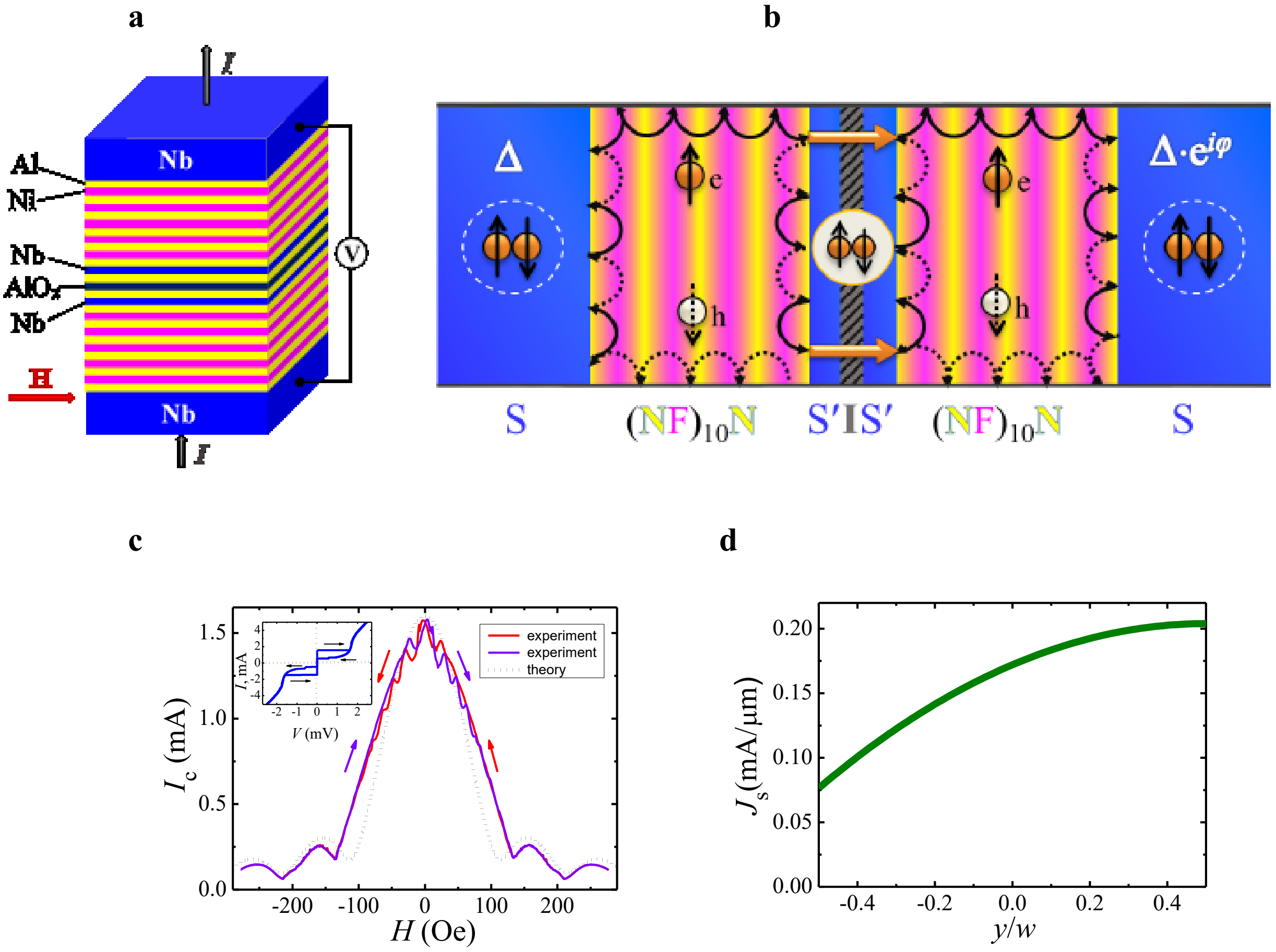

**Fig. 2 S(NF)$_{10}$NS′IS′(NF)$_{10}$NS multilayer.** **a**, Schematic view of a 10 μm × 10 μm S(NF)$_n$NS′IS′(NF)$_n$NS Josephson junction and its biasing. A device with $n$ = 5 is shown for simplicity. **b**, Semi-classical skipping orbit image of charge trajectories in the *yz*-plane illustrating the transformation of quasiparticle trajectories, confined to its side surfaces of the (NF)$_{10}$N multilayer, into much more uniform supercurrent flow across the inner S′IS′ trilayer. **c**, Main panel: $I_c(H)$ patterns measured for two directions of field sweeping (solid lines). The black dots show a theoretical $I_c(H)$ dependence calculated using the supercurrent density profile displayed in Fig. 2d. The inset shows the measured *I-V* curve. **d**, Extracted spatial distribution of the supercurrent density profile $J_s(y)$ through the S′IS′ interior trilayer integrated over the $x$ direction. The measurement temperature was 4.2 K.

The coincidence of the oscillation period with the value of the magnetic flux quantum $\Phi_0$ shows that the supercurrent through the inner S′IS′trilayer is distributed much more uniformly in space than in the multilayer (NF)$_{10}$N structures. Some nonuniformity may be associated with the fact that lateral size of the JJ, 10 μm, is larger than the Josephson penetration depth, $\lambda_J$ (estimated to be about 4 μm for the S′IS′ JJ). The presence of (NF)$_{10}$N structures is manifested in the low-amplitude "ripples" seen in the measured $I_c(H)$ characteristic. The reconstructed spatial profile of the supercurrent flowing through the S′IS′ trilayer and controlling the $I_c(H)$ pattern of the entire junction is displayed in Fig. 2d.

### D. S(NF)$_n$NS′IS junctions

To further confirm the idea that after passing via the edges of the (NF)$_n$N multilayered structure, the supercurrent spreads much more evenly through the S′ superconducting layer, we have fabricated and studied Nb(120)[Al(3.1)Ni(1.2)]$_{10}$Al(3.1)Nb(12)Al/AlO$_x$(9.3)Al(3.1)Nb(480) [S(NF)$_{10}$NS′IS] devices. As above, the S′IS trilayer was again a conventional Nb/Al/AlO$_x$/Al/Nb Josephson junction with a 12-nm thick Nb(S′) film (Figs. 3a and 3b). Its hysteretic *I-V* characteristic is shown in the inset in Fig. 3c, while the main panel of Fig. 3c demonstrates the measured Fraunhofer-like $I_c(H)$ dependence.

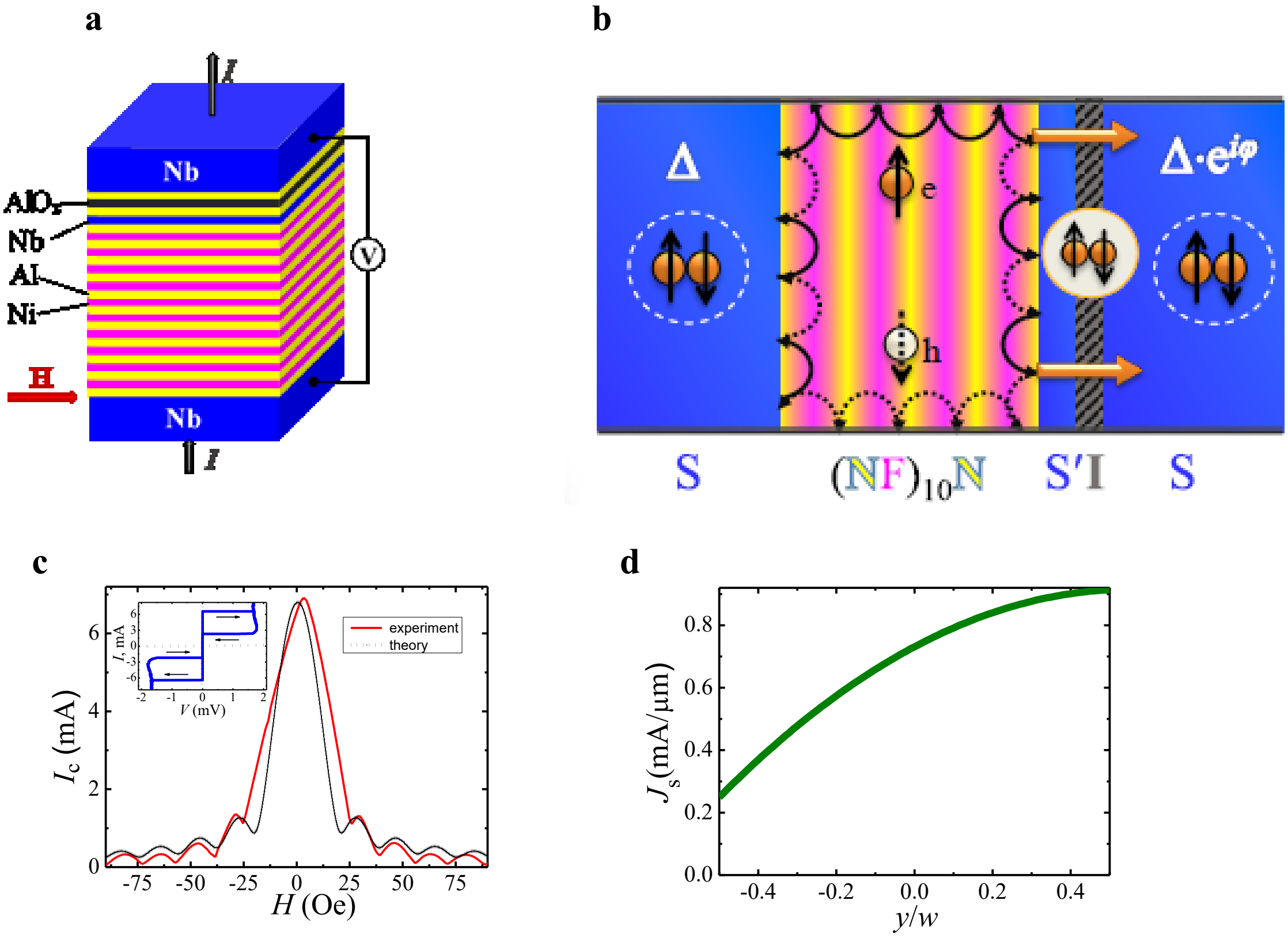

**Fig. 3 S(NF)$_{10}$NS′IS multilayers. a**, Schematic view of a 10 µm × 10 µm S(NF)$_n$NS′IS Josephson junction and its biasing. **b**, Semi-classical skipping orbit image illustrating the transformation of charge trajectories in the *yz*-plane illustrating the transformation of quasiparticle trajectories, confined to its side surfaces of the (NF)$_{10}$N multilayer, into a more uniform supercurrent flow across the S′IS′ trilayer. **c**, Main panel: a measured pattern (red solid line) and the theoretical (black dots) $I_c(H)$ dependence calculated using the supercurrent density profile displayed in Fig. 3d. The inset shows the measured *I-V* curve. **d**, Extracted spatial distribution of the supercurrent density profile $J_s(y)$ through the S′IS trilayer integrated over the *x* direction. The measurement temperature was 4.2 K.

As in the former case of S(NF)$_{10}$NS′IS′(NF)$_{10}$NS devices, the impact of the (NF)$_{10}$N multilayer magnetization on the $I_c(H)$ dependence for the S′IS junction was expected to be minimal. Indeed, a small stray field manifested itself in a slight shift of the $I_c(H)$ dependence along the $H$ axis, visible in Fig. 3c. Taking as above $\mu = 1$, we get the sensing area of the S′IS junction $s = 1.1$ μm$^2$. The period of the Fraunhofer pattern following from Fig. 3c $\Delta H \approx 18.8$ Oe gives the respective magnetic flux period $\Delta\Phi = 2.1\cdot10^{-7}$ G·cm$^2$ in perfect agreement with the magnetic flux quantum value $\Phi_0 = 2.068\cdot10^{-7}$ G·cm$^2$. The reconstructed spatial profile of the supercurrent through the S′IS trilayer, which controls the $I_c(H)$ characteristic of the entire device, is displayed in Fig. 3d. In this case, the dependence of $J_s(y)$ on the coordinate is more significant than in Fig. 2d, which results in the measured $I_c(H)$ curve appearing to be a less ideal Fraunhofer pattern and more typical for distributed JJs (compare the measured and calculated data in Fig. 3c). Indeed, in this case, the estimated value $\lambda_J = 2.4$ μm is considerably smaller than the lateral size of the JJ, 10 μm.

## IV. DISCUSSION

Engineering new states of matter using multilayer technology makes it possible to create previously unattainable hybrid states and ordered phases required for quantum technologies [30, 31]. As argued in the paper [21], our multilayered barrier at 4.2 K is in a *superparamagnetic* state slightly above the blocking temperature. In this case, magnetic moments of the magnetic clusters are switching between the two directions of the easy axis, locally destroying time-reversal symmetry similar to a topological photonic alloy with a mixture of non-magnetized and magnetized components [32]. In the photonic analog, the presence of substitutional disorder led to the emergence of topologically non-trivial states via a topological invariant determined by the appearance of a 2$\pi$ reflection phase change in the scattered waves [32]. Even more, chiral edge states can be created without the need to break time-reversal symmetry everywhere inside the sample since a local breakdown of time-reversal symmetry is sufficient for this purpose. Therefore, by analogy with the findings [32], we can expect the manifestation of similar properties in our Josephson junctions where hybrid $(NF)_nN$ multilayered barriers with a local breakdown of time-reversal symmetry provide a fundamental mechanism for controlling the basic properties of the supercurrent flow across them.

Let us analyze the presence of the above specific properties in our samples and discuss the role of constituent elements of the complex heterostructures in the creation and manipulation of robust edge states exhibiting chirality features. (1) *Edge localized current-carrying modes* manifesting themselves in the SQUID-like characteristics, Fig. 1d. In our previous publication [21], the emergence of the edge states was attributed to dephasing accumulated by an electron (or a hole) during its chaotic (while maintaining coherence) motion across the strongly *inhomogeneous* and *anisotropic* weak link, which leads to strong spatial decay of the bulk contribution and the predominance of near-surface trajectories; see also the paper [33]. It is clear that the survival effect is the more successful, the smaller is the phase space over which the electron trajectories are averaged. Therefore, the contribution of *hinge* states to the total current in comparison with other near-boundary transport channels may be decisive. (2) *Chirality of the current-carrying modes*, revealing itself in periodicity with a normal ($hc/e$) rather than a superconducting ($hc/2e$) flux quantum (Fig. 1c), and the transformation of edge currents into hybrid electron-hole modes propagating along the contacting superconducting electrode (Figs. 2c and 3c). The chiral-like transfer o Cooper pairs occurs due to time-reversal symmetry breaking caused by the intrinsic magnetization of nm-thick Ni layers and enhanced with the multilayer structure [34]. (3) *Robust* and *phase-coherent transport of superconducting correlations* arising in $(NF)_nN$ multilayers despite a large number of electronically mismatched interfaces, possible disorder inside the films, and the presence of an insulating interlayer can be a consequence of topological protection similar to related effects in photonic alloys [32]. Note that in this case, the emergence of a topological state does not require strong spin-orbit coupling and/or external magnetic fields.

Finally, chiral and robust propagation of charge carriers in three-dimensional materials remains a highly sought-after task in quantum technologies, and to solve it, one needs easily accessible and well-controlled materials, as well as new, reliable platforms. The key novelty of our work is the demonstration of the possibility of creating desired materials capable of harboring protected edge current-carrying modes by stacking alternatively nonmagnetic and ferromagnetic ultrathin layers. Such engineering methodology based on conventional and well-studied metals inspires optimism regarding its potential as a new tunable platform for exploring the non-trivial edge physics and achieving dissipation-free transport in the devices of next-generation electronics [35]. The multilayer approach proposed in the work fits well with the new 3D heterogeneous integration technology, which is based on the vertical stacking of ultrathin layers of various types, thereby creating novel 3D circuit architectures with high integration density [36, 37].

## V. CONCLUSIONS

Our study presents experimental evidence of edge-confined, and potentially chiral, current-carrying modes in metallic multilayers composed of alternating nanometer-scale Al and Ni films. Using Josephson interferometry, we observed interference patterns in the maximum critical current versus in-plane magnetic field characteristics, which radically deviate from the conventional Fraunhofer form and exhibit features consistent with SQUID-like periodicity and period-doubling, the hallmarks of chiral Andreev edge transport. These effects become pronounced as the number of Al/Ni bilayers increases, suggesting a collective origin associated with the multilayer architecture.

While the observed phenomena bear a striking resemblance to those reported in topological systems with higher-order edge modes, our heterostructures are built from ordinary metals with no inherent topological band structure. This raises intriguing questions about the possible emergence of topologically-like behavior in synthetic structures through mechanisms such as interface-induced symmetry breaking, magnetization inhomogeneity, and/or proximity effects. Although the definitive topological interpretation requires further validation, our results point to a previously unrecognized pathway for engineering exotic edge transport in multilayered metallic systems. These findings open up new directions for realizing low-dissipation quantum devices using technologically accessible and structurally tunable materials.

## ACKNOWLEDGMENTS

I.P.N. acknowledges financial support from NSF grant DMR 1905742 and from the NSF DISCoVER Expedition award under grant no. CCF-2124453. I.P.N. is also grateful for the opportunity to use the facilities of the Materials Research Center at Northwestern University, supervised by J. B. Ketterson and supported by NSF. M.B. acknowledges support of the EU NextGenerationEU through the Recovery and Resilience Plan for Slovakia under Project 09I03-03-V01-00139. The authors acknowledge using the results of HRTEM analysis of SFT devices carried out by Dr. N. Missert during the C3 Program.

## REFERENCES

[1] E. Prada, P. San-Jose, M. W. A. de Moor, A. Geresdi, E. J. H. Lee, J. Klinovaja, D. Loss, J. Nygård, R. Aguado, and L. P. Kouwenhoven, From Andreev to Majorana bound states in hybrid superconductor–semiconductor nanowires, Nat. Rev. Phys. 2, 575 (2020). [https://doi.org/10.1038/s42254-020-0228-y]

[2] A. Barone and G. Paternò, Physics and Applications of the Josephson Effect (Wiley, New York, 1982).

[3] R. Gross, A. Marx, and F. Deppe, Applied Superconductivity: Josephson Effect and Superconducting Electronics (De Gruyter, Berlin, 2016).

[4] R. Citro, C. Guarcello, and S. Pagano, Josephson junctions, superconducting circuits, and qubits for quantum technologies, in New Trends and Platforms for Quantum Technologies, edited by R. Aguado, R. Citro, M. Lewenstein, and M. Stern (Springer, Cham, 2024), p. 1. [https://doi.org/10.1007/978-3-031-55657-9_1]

[5] J. Qi, C.-Z. Chen, J. Song, J. Liu, K. He, Q.-F. Sun, and X. C. Xie, Edge supercurrent in Josephson junctions based on topological materials, Sci. China Phys. Mech. Astron. 68, 227401 (2025). [https://doi.org/10.1007/s11433-024-2520-9]

[6] N. Mason, Superconductivity on the edge, Science 352, 891 (2016). [https://doi.org/10.1126/science.aaf6604]

[7] R. Moessner and A. P. Mackenzie, Topological Phases of Matter (Cambridge University Press, Cambridge, 2021). [https://doi.org/10.1017/9781316226308]

[8] A. Anirban, 15 years of topological insulators, Nat. Rev. Phys. 5, 267 (2023). [https://doi.org/10.1038/s42254-023-00587-y]

[9] Y. Ando and L. Fu, Topological crystalline insulators and topological superconductors: From concepts to materials, Annu. Rev. Condens. Matter Phys. 6, 361 (2015). [https://doi.org/10.1146/annurev-conmatphys-031214-014501]

[10] B. Liu and W. Zhang, Research progress of topological quantum materials: From first-order to higher-order, Symmetry 15, 1651 (2023). [https://doi.org/10.3390/sym15091651]

[11] O. Breunig and Y. Ando, Opportunities in topological insulator devices, Nat. Rev. Phys. 4, 184 (2022). [https://doi.org/10.1038/s42254-021-00402-6]

[12] Y.-B. Yang, J.-H. Wang, K. Li, and Y. Xu, Higher-order topological phases in crystalline and non-crystalline systems: A review, J. Phys. Condens. Matter 36, 283002 (2024). [https://doi.org/10.1088/1361-648X/ad3abd]

[13] Y. Shen, Z. Li, Q. Niu, and Z. Qiao, Disorder-induced phase transitions in three-dimensional chiral second-order topological insulator, Phys. Rev. B 109, 035303 (2024). [https://doi.org/10.1103/PhysRevB.109.035303]

[14] R. Noguchi, T. Takahashi, K. Kuroda, T. Ochi, T. Shirasawa, M. Sakano, C. Bareille, M. Nakayama, M. D. Watson, K. Yaji, A. Harasawa, H. Iwasawa, P. Dudin, T. K. Kim, M. Hoesch, V. Kandyba, A. Barinov, T. Sasagawa, T. Kondo, and T. Sato, Evidence for a higher-order topological insulator in a three-dimensional material built from van der Waals stacking of

bismuth-halide chains, Nat. Mater. 20, 473 (2021). [https://doi.org/10.1038/s41563-020-00859-2]

[15] Z. Xiong, R.-Y. Zhang, R. Yu, C. T. Chan, and Y. Chen, Hidden symmetry-enforced nexus points of nodal lines in layer-stacked dielectric photonic crystals, Light Sci. Appl. 9, 176 (2020). [https://doi.org/10.1038/s41377-020-00403-0]

[16] Z. Wang, D. Liu, H. T. Teo, Q. Wang, H. Xue, and B. Zhang, Higher-order Dirac semimetal in a photonic crystal, Phys. Rev. B 105, L060101 (2022). [https://doi.org/10.1103/PhysRevB.105.L060101]

[17] Y. Zhang, J. Tang, X. Dai, S. Zhang, and Y. Xiang, Higher-order nodal ring photonic semimetal, Opt. Lett. 47, 5885 (2022). [https://doi.org/10.1364/OL.475147]

[18] H.-R. Xia, J.-Z. Li, S.-Y. Yuan, and M. Xiao, Photonic realization of chiral hinge states in a Chern-insulator stack, Phys. Rev. B 111, L041112 (2025). [https://doi.org/10.1103/PhysRevB.111.L041112]

[19] J.-H. Wang, Y.-B. Yang, N. Dai, and Y. Xu, Structural-disorder-induced second-order topological insulators in three dimensions, Phys. Rev. Lett. 126, 206404 (2021). [https://doi.org/10.1103/PhysRevLett.126.206404]

[20] H.-G. Meyer, L. Fritzsch, S. Anders, M. Schmelz, J. Kunert, and G. Oelsner, Josephson junctions, in Applied Superconductivity: Handbook on Devices and Applications, edited by P. Seidel (Wiley-VCH, Weinheim, 2015), Vol. 1, p. 281.

[21] I. P. Nevirkovets, M. A. Belogolovskii, and J. B. Ketterson, Josephson junctions with artificial superparamagnetic barrier: A promising avenue for nanoscale magnetometry, Phys. Rev. Appl. 14, 014092 (2020). [https://doi.org/10.1103/PhysRevApplied.14.014092]

[22] A. Mani and C. Benjamin, Probing helicity and the topological origins of helicity via non-local Hanbury-Brown and Twiss correlations, Sci. Rep. 7, 6954 (2017). [https://doi.org/10.1038/s41598-017-06820-w]

[23] H. Knüpfer, C. B. Muratov, and F. Nolte, Magnetic domains in thin ferromagnetic films with strong perpendicular anisotropy, Arch. Ration. Mech. Anal. 232, 727 (2019). [https://doi.org/10.1007/s00205-018-1332-3]

[24] Z. B. Guo, W. B. Mi, Q. Zhang, R. Zhang, O. Aboljadayel, and X. X. Zhang, Anomalous Hall effect in polycrystalline Ni films, Solid State Commun. 152, 220 (2012). [https://doi.org/10.1016/j.ssc.2011.10.039]

[25] C. Delacour, L. Ortega, M. Faucher, T. Crozes, T. Fournier, B. Pannetier, and V. Bouchiat, Persistence of superconductivity in niobium ultrathin films grown on R-plane sapphire, Phys. Rev. B 83, 144504 (2011). [https://doi.org/10.1103/PhysRevB.83.144504]

[26] I. P. Nevirkovets, A superconducting transistor with improved isolation between the input and output terminals, Supercond. Sci. Technol. 22, 105009 (2009). [https://doi.org/10.1088/0953-2048/22/10/105009]

[27] R. C. Dynes and T. A. Fulton, Supercurrent density distribution in Josephson junctions, Phys. Rev. B 3, 3015 (1971). [https://doi.org/10.1103/PhysRevB.3.3015]

[28] M. Weihnacht, Influence of film thickness on D.C. Josephson current, Phys. Status Solidi B 32, K169 (1969). [https://doi.org/10.1002/pssb.19690320259]

[29] G. Wild, C. Probst, A. Marx, and R. Gross, Josephson coupling and Fiske dynamics in ferromagnetic tunnel junctions, Eur. Phys. J. B 78, 509 (2010). [https://doi.org/10.1140/epjb/e2010-10636-4]

[30] A. R.-P. Montblanch, M. Barbone, I. Aharonovich, M. Atatüre, and A. C. Ferrari, Layered materials as a platform for quantum technologies, Nat. Nanotechnol. 18, 555 (2023). [https://doi.org/10.1038/s41565-023-01354-x]

[31] S. Li, M. Gong, S. Cheng, H. Jiang, and X. C. Xie, Dissipationless layertronics in axion insulator $MnBi_2Te_4$, Natl. Sci. Rev. 11, nwad262 (2023). [https://doi.org/10.1093/nsr/nwad262]

[32] T. Qu, M.Wang, X. Cheng, X. Cui, R.-Y. Zhang, Z.-Q. Zhang, L. Zhang, J. Chen, and C. T. Chan, Topological photonic alloy, Phys. Rev. Lett. 132, 223802 (2024). [https://doi.org/10.1103/PhysRevLett.132.223802]

[33] I. L. Aleiner, A. V. Andreev, and V. M. Vinokur, Aharonov-Bohm oscillations in singly connected disordered conductors, Phys. Rev. Lett. 114, 076802 (2015). [https://doi.org/10.1103/PhysRevLett.114.076802]

[34] X. Lu, J. Zou, J. M. Pham, A. A. Rana, C.-T. Liao, E. C. Subramanian, X. Wu, Y. H. Lo, C. S. Bevis, R. M. Karl, Jr., S. Lepadatu, Y.-S. Yu, Y. Tserkovnyak, T. P. Russell, D. A. Shapiro, H. C. Kapteyn, M. M. Murnane, R. Streubel, and J. Miao, Visualizing magnetic order in self-assembly of superparamagnetic nanoparticles, arXiv:2401.01284 [cond-mat.mes-hall] (2024). [https://doi.org/10.48550/arXiv.2401.01284]

[35] M. J. Gilbert, Topological electronics, Commun. Phys. 4, 70 (2021). [https://doi.org/10.1038/s42005-021-00569-5]

[36] S. Reda, 3D integration advances computing, Nature 547, 38 (2017). [https://doi.org/10.1038/547038a]

[37] J. H. Kang, J. Choe, D. Kim, J. Lee, G. Lee, H. Kim, Y.-M. Kim, E. Choi, B. H. Park, and H. S. Lee, Monolithic 3D integration of 2D materials-based electronics towards ultimate edge computing solutions, Nat. Mater. 22, 1470 (2023). [https://doi.org/10.1038/s41563-023-01704-z]

[38] A. Sáenz-Trevizo and A. M. Hodge, Nanomaterials by design: A review of nanoscale metallic multilayers, Nanotechnology 31, 292002 (2020). [https://doi.org/10.1088/1361-6528/ab7b2a]

[39] C. Rizal, B. Moa, and B. B. Niraula, Ferromagnetic multilayers: Magnetoresistance, magnetic anisotropy, and beyond, Magnetochemistry 2, 22 (2016). [https://doi.org/10.3390/magnetochemistry2020022]

[40] M. Wang, L. Qiu, X. Zhao, Y. Li, T. Rao, S. He, L. Qin, and J. Tao, Multilayered Al/Ni energetic structural materials with high energy density and mechanical properties prepared by a facile approach of electrodeposition and hot pressing, Mater. Sci. Eng. A 757, 23 (2019). [https://doi.org/10.1016/j.msea.2019.04.074]

[41] S. S. Riegler, Y. H. S. Camposano, K. Jaekel, M. Frey, C. Neemann, S. Matthes, E. Vardo, M. R. Chegeni, H. Bartsch, R. Busch, J. Müller, P. Schaaf, and I. Gallino, Nanocalorimetry of

nanoscaled Ni/Al multilayer films: On the methodology to determine reaction kinetics for highly reactive films, Adv. Eng. Mater. 26, 2302279 (2024). [https://doi.org/10.1002/adem.202302279]

[42] K. Le Guen, G. Gamblin, P. Jonnard, M. Salou, J. Ben Youssef, S. Rioual, and B. Rouvellou, Spectroscopic study of interfaces in Al/Ni periodic multilayers, Eur. Phys. J. Appl. Phys. 45, 20502 (2009). [https://doi.org/10.1051/epjap/2009027]

[43] N. Kaplan, H. Kuru, and H. Köçkar, Investigation of the influence of Al layer and total film thicknesses on structural and related magnetic properties in sputtered Ni/Al multilayer thin films, J. Mater. Sci. Mater. Electron. 35, 302 (2024). [https://doi.org/10.1007/s10854-023-11940-0]

[44] A. Karpuz, H. Köçkar, and S. Çolmekçi, Structural and corresponding magnetic properties of sputtered Ni/Al multilayer films: Effect of Ni layer thickness, Acta Phys. Pol. A 134, 1180 (2018). [https://doi.org/10.12693/APhysPolA.134.1180]

[45] I. P. Nevirkovets and O. A. Mukhanov, Memory cell for high-density arrays based on a multiterminal superconducting-ferromagnetic device, Phys. Rev. Appl. 10, 034013 (2018). [https://doi.org/10.1103/PhysRevApplied.10.034013]

[46] R. P. Cowburn, Property variation with shape in magnetic nanoelements, J. Phys. D Appl. Phys. 33, R1 (2000). [https://doi.org/10.1088/0022-3727/33/1/201]

[47] C. A. Neugebauer, Saturation magnetization of nickel films of thickness less than 100 Å, Phys. Rev. 116, 1441 (1959). [https://doi.org/10.1103/PhysRev.116.1441]

[48] I. P. Nevirkovets and O. A. Mukhanov, Peculiar interference pattern of Josephson junctions involving periodic ferromagnet-normal metal structure, Supercond. Sci. Technol. 31, 03LT01 (2018). [https://doi.org/10.1088/1361-6668/aaa9f8]

[49] S. Bedanta and W. Kleemann, Supermagnetism, J. Phys. D Appl. Phys. 42, 013001 (2009). [https://doi.org/10.1088/0022-3727/42/1/013001]

[50] A. Sharma, N. Theodoropoulou, T. Haillard, R. Acharyya, R. Loloee, W. P. Pratt, Jr., J. Zhang, and M. A. Crimp, Current-perpendicular-to-plane magnetoresistance of ferromagnetic F/Al interfaces (F = Py, Co, Fe, and $Co_{91}Fe_{9}$) and structural studies of Co/Al and Py/Al, Phys. Rev. B 77, 224438 (2008). [https://doi.org/10.1103/PhysRevB.77.224438]

[51] Y. Wang, Z. Xing, Y. Qiao, H. Jiang, X. Yu, F. Ye, Y. Li, L. Wang, and B. Liu, Asymmetric atomic diffusion and phase growth at the Al/Ni and Ni/Al interfaces in the Al-Ni multilayers obtained by magnetron deposition, J. Alloys Compd. 789, 887 (2019). [https://doi.org/10.1016/j.jallcom.2019.03.104]

[52] A. I. Braginski, Superconductor electronics: Status and outlook, J. Supercond. Nov. Magn. 32, 23 (2019). [https://doi.org/10.1007/s10948-018-4884-7]

[53] K. R. Joshi, S. Ghimire, M. A. Tanatar, A. Datta, J.-S. Oh, L. Zhou, C. J. Kopas, J. Marshall, J. Y. Mutus, J. Slaughter, M. J. Kramer, J. A. Sauls, and R. Prozorov, Quasiparticle spectroscopy, transport, and magnetic properties of Nb films used in superconducting qubits, Phys. Rev. Appl. 20, 024031 (2023). [https://doi.org/10.1103/PhysRevApplied.20.024031]

[54] K. M. Ryan, C. G. Torres-Castanedo, D. P. Goronzy, D. A. G. Wetten, M. Field, C. J. Kopas, J. Marshall, M. J. Reagor, M. J. Bedzyk, M. C. Hersam, and V. Chandrasekhar, Characterization of Nb films for superconducting qubits using phase boundary measurements, Appl. Phys. Lett. 121, 202601 (2022). [https://doi.org/10.1063/5.0121410]

[55] R. M. L. McFadden, M. Asaduzzaman, T. Prokscha, Z. Salman, A. Suter, and T. Junginger, Depth-resolved measurements of the Meissner screening profile in surface-treated Nb, Phys. Rev. Appl. 19, 044018 (2023). [https://doi.org/10.1103/PhysRevApplied.19.044018]

[56] R. M. L. McFadden and T. Junginger, Search for inhomogeneous Meissner screening in Nb induced by low-temperature surface treatments, AIP Adv. 14, 095320 (2024). [https://doi.org/10.1063/5.0226022]

[57] T. S. Khaire, W. P. Pratt, Jr., and N. O. Birge, Critical current behavior in Josephson junctions with the weak ferromagnet PdNi, Phys. Rev. B 79, 094523 (2009). [https://doi.org/10.1103/PhysRevB.79.094523]

[58] V. Niedzielski, E. C. Gingrich, R. Loloee, W. P. Pratt, Jr., and N. O. Birge, S/F/S Josephson junctions with single-domain ferromagnets for memory applications, Supercond. Sci. Technol. 28, 085012 (2015). [https://doi.org/10.1088/0953-2048/28/8/085012]

[59] J. M. Rowell, Magnetic field dependence of the Josephson tunnel current, Phys. Rev. Lett. 11, 200 (1963). [https://doi.org/10.1103/PhysRevLett.11.200]

[60] N. O. Birge and N. Satchell, Ferromagnetic materials for Josephson π junctions, APL Mater. 12, 041105 (2024). [https://doi.org/10.1063/5.0200544]

[61] I. P. Nevirkovets, Observation of fractional vortices and π phases in Josephson junctions involving periodic magnetic layers, Phys. Rev. B 108, 024503 (2023). [https://doi.org/10.1103/PhysRevB.108.024503]

[62] I. P. Nevirkovets, M. A. Belogolovskii, O. A. Mukhanov, and J. B. Ketterson, Magnetic field sensor based on a single Josephson junction with a multilayer ferromagnet/normal metal barrier, IEEE Trans. Appl. Supercond. 31, 1800205 (2021). [https://doi.org/10.1109/TASC.2021.3050224]

[63] A. W. Kleinsasser, High performance Nb Josephson devices for petaflops computing, IEEE Trans. Appl. Supercond. 11, 1043 (2001). [https://doi.org/10.1109/77.919545]

[64] X. He, W. Zhong, C.-T. Au, and Y. Du, Size dependence of the magnetic properties of Ni nanoparticles prepared by thermal decomposition method, Nanoscale Res. Lett. 8, 446 (2013). [https://doi.org/10.1186/1556-276X-8-446]

# APPENDICES

## Appendix A: Magnetic and electrical characteristics of nanoscale heterostructures formed by alternating Al/Ni bilayers

Nanoscale metallic multilayers based on alternating layers of various conductors are exhibiting a wide range of exceptional properties associated with a large number of interfaces and the nanometer layer thicknesses, and therefore significantly different from those observed in related single films [38]. In particular, it relates stacks of nm-thick non-magnetic/ferromagnetic (N/F) bilayers with unusual behavior caused by magnetic anisotropy associated with the N/F interfaces and magnetic coupling through the N spacer [39]. Among them, metallic multilayers based on Al/F pairs have attracted great interest due to their highly energetic nature [40, 41]. Due to the limited number of stable intermetallic compounds (nickel aluminides) in the corresponding phase diagram [42] and other unique properties, the couple of Al and Ni has proven to be in demand for various applications [43].

In the following, we will limit ourselves to discussing the magnetic properties of Al/Ni multilayers and their dependences on the Al [43] and Ni [44] thicknesses, $d_{Al}$ and $d_{Ni}$. Although $d_{Al}$ and $d_{Ni}$ values in the papers [43, 44] were significantly larger than in our samples, we can use their results at least to estimate the expected magnetic parameters for our Al/Ni multilayers. The closest data, although quite far from our parameters, can be found in the paper [44] for a quite large-area heterostructure formed by eight Al/Ni bilayers with equal thicknesses $d_{Al} = d_{Ni} = 10$ nm. It exhibited the coercivity $H_c = 30$ Oe [44] and the specific saturation magnetization $M_s = 272$ emu/cm$^3$ [44] that agrees well with $M_s = 290$ emu/cm$^3$ for our 4.7 nm thick Ni film [45]. Note that the $M_s$ and $H_c$ values are significantly affected by variations of the multilayer content due to the changes in the growth processes [43] and are expected to be strongly size-sensitive [46].

To show the quality of our Al/F multilayers, Fig. A1 demonstrates a TEM (transmission electron microscopy) image of the cross-section of a fragment of a representative Nb(120)$[Al(3.1)Ni(1.2)]_{10}$Al/AlO$_x$(3.1)$[Al(3.1)Ni(1.2)]_{10}$Al(3.1)Nb(75) Josephson junction; the numbers in the parentheses are nominal thicknesses of the respective layers in nm. Individual Al and Ni layers are well visible within the periodic S$(N/F)_{10}$NI$(N/F)_{10}$NS structure. However, the quality of the image does not allow concluding whether or not 1.2 nm-thick Ni films are continuous or rather an array of small, superparamagnetic particles with an average volume of a few nm$^3$ and the concentration about $10^{12}$ cm$^{-2}$, as was found in Ref. [47] for Ni films.

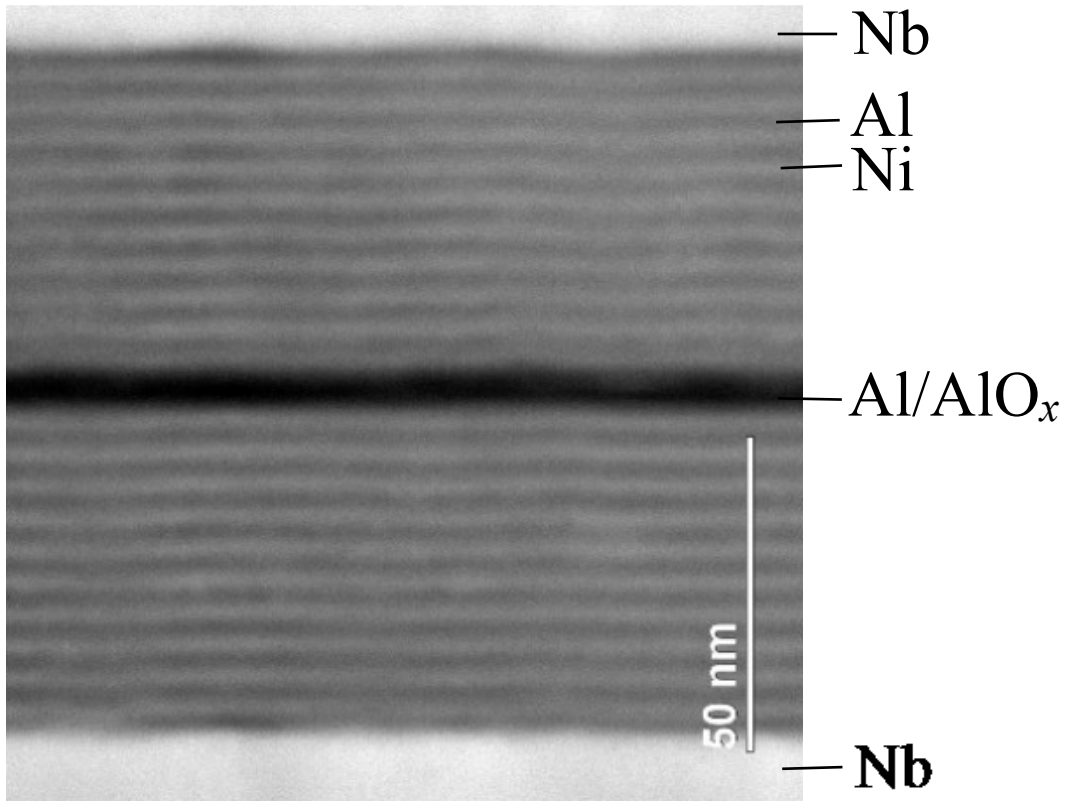

**Fig. A1**. TEM image of the cross-section of a fragment of $S(N/F)_{10}NI(N/F)_{10}NS$ Josephson junction, where lighter and darker layers are Ni and Al, respectively.

To characterize the magnetic state of our Al/Ni multilayers with fixed Al and Ni thicknesses $d_{Al}$ = 3.10 nm and $d_{Ni}$ = 1.34 nm we fabricated a large area (5.5 mm × 11.4 mm, *i.e.*, 0.63 cm$^2$) Nb/(Al/Ni)$_{70}$ sample on a Si/SiO$_2$ substrate [48]. At 10 K, just above the critical temperature of Nb $T_c \approx 9$ K, the magnetic moment $M$ as a function of the magnetic field $H$ parallel to the layers was saturated at $H \approx 2$ kOe with $M_{sat}$ = 1.15 10$^{-4}$ emu for the (Al/Ni)$_{70}$ periodic structure of the volume of 1.95 10$^{-5}$ cm$^3$. Hence, in our samples, the specific saturation magnetization $M_s$ = 5.9 emu/cm$^3$ was almost two orders of magnitude smaller than the corresponding value for the (Al/Ni)$_8$ heterostructure with $d_{Al} = d_{Ni}$ = 10 nm [44], see above. At the same time, the total thickness of the nickel films was almost the same (94 nm in our Nb/(Al/Ni)$_{70}$ sample and 80 nm in the (Al/Ni)$_8$ multilayer). The difference arises due to the significant surface-to-bulk ratio in thin films, where the magnetization is usually reduced compared to the bulk materials. For this reason, the magnetic moment of the thicker films in the work [44] is closer to that of the bulk, whereas our much thinner films have significantly smaller magnetic moments. This effect is enhanced further by the influence of the substrate interface, where the crystal structure and magnetic properties can differ significantly from those of the bulk material.

The analysis for the multilayer that is comprised of 70 Al/Ni periods [21] has shown that our working temperature $T$ = 4.2 K is slightly above the blocking temperature. For small $H$, its $M$-vs-$H$ dependence is a linear function [49] whose slope determines the magnetic susceptibility $\chi$ = 0.33. The value of permeability $\mu = 1 + 4\pi\chi = 5.1$ is in agreement with the data of our direct measurements, $\mu$ = 5.4 [21]. In the main text of the paper and below, we calculate characteristic magnetic lengths of related Josephson junctions using the value $\mu \approx 5$. It was determined for multilayers with a very large number of bilayers $n$ = 70 and for a very large area $S$ = 0.63 cm$^2$. Therefore, for each sample analyzed below, we will use two limiting values $\mu$ = 1 and $\mu$ = 5 and consider the one that gives the best fitting with the corresponding experiment. At the same time, measuring the magnetic properties of the multilayers remains the most important task of subsequent experiments.

As for charge transport through $(Al/Ni)_n$ multilayers, a significant factor limiting the transparency of such heterostructures is interface scattering, the source of which lies in the mismatch between the Fermi velocities of two different materials. In the bilayers composed of a non-magnetic N and spin-polarized F metals, such scattering differs for electrons with moments along (↑) the direction of the local magnetization **M** or opposite (↓) to **M**. Accordingly, there are two specific resistances, $AR^{\uparrow}_{\mathrm{N/F}}$ and $AR^{\downarrow}_{\mathrm{N/F}}$ ($A$ is the area though which the current flows), which define the interface specific resistance $2AR^{*}_{\mathrm{N/F}} = AR^{\uparrow}_{\mathrm{N/F}} + AR^{\downarrow}_{\mathrm{N/F}}$, and the scattering asymmetry $\gamma_{\mathrm{N/F}} = (AR^{\uparrow}_{\mathrm{N/F}} - AR^{\downarrow}_{\mathrm{N/F}})/(AR^{\uparrow}_{\mathrm{N/F}} + AR^{\downarrow}_{\mathrm{N/F}})$. The authors of Ref. [50] determined these parameters for Al/F interfaces (F: Py = $Ni_{84}Fe_{16}$, Co, Fe, and $Co_{91}Fe_{9}$) and found large values of $2AR^{*}_{\mathrm{N/F}} = 8.4 - 11.6\ \mathrm{f\Omega \cdot m^2}$ for all four cases while $\gamma_{\mathrm{N/F}}$ had the lowest value of 0.025 just in the Al/Py bilayers. Let us also note that due to the violation of the electronic band structure, asymmetric atomic diffusion and phase growth at the non-magnet – ferromagnet interface can lead to the emergence of interfacial magnetic anisotropy, in particular, observed in magnetron sputtered Al/Ni multilayers when Al atoms were deposited on Ni or, conversely, Ni atoms were deposited on Al [51].

## Appendix B. London penetration depth in Nb films

Niobium, a conventional (*s*-band) low-temperature superconductor with a critical temperature about 9 K is the material of choice for various superconducting applications [52]. Since its discovery in 1930s, it has become one of the most studied S metals. Nevertheless, there are still some issues requiring clarification and further study. These include, in particular, spectroscopic methods using analysis based on the London penetration depth $\lambda_L$ [53]. There are great differences between bulk niobium and Nb thin films in terms of morphology, purity, and relative length scales [54]. For this reason, active research into niobium films, driven primarily by their use in superconducting qubits, continues [54]. For sufficiently thick Nb samples, two spatial regions, one adjacent to the surface and one deep below the surface, were found. The second area, with the Meissner response close to that in the bulk with $\lambda_{\mathrm{bulk}} = 29$ nm [55], starts from depths exceeding 120 nm [56]. At smaller depths, the magnetic-field penetration is described by the parameter $\lambda_{\mathrm{surface}} > \lambda_{\mathrm{bulk}}$ and in the near-surface region (at a distance from the surface of less than 40 nm) $\lambda_{\mathrm{surface}}$ becomes spatially dependent on depth [56]. For most of the niobium samples studied in the work [56], the near-surface penetration depth of the magnetic field was slightly less than 80 nm. In the works [57,58], the value of $\lambda$ for Nb films serving as electrodes in the Josephson junction was fixed at 85 nm.

A brief literature review of the $\lambda_{Nb}$ values indicates that it should be found for each specific technology of the Josephson junction fabrication. That is why below we first determine $\lambda_{Nb}$ using the dependence of the maximum supercurrent $I_c$, versus in-plane magnetic field $H$ for the conventional SNINS junction (S = Nb), where $\lambda_{Nb}$ and the $I_c(H = 0)$ amplitude are two fitting parameters, see the subsection D.1. After that, the $\lambda_{Nb}$ value found in this way is used to calculate the effective magnetic thickness $t$ in multilayered samples with a sufficiently large number of (F/N) bilayers, where the contribution of the related weak link is dominant while variations in the $\lambda_{Nb}$ value in the S electrodes do not play a significant role.

## Appendix C. Maximum supercurrent - versus - in-plane magnetic field interference pattern in Josephson junctions

Seminal works [2, 59, 27] showed that the distribution of supercurrent in a two-terminal Josephson junction can be probed by analyzing an interference pattern, *i.e.*, the function of the maximum supercurrent $I_c$ from the in-plane magnetic field $H$. As long as the supercurrent density $J_s$ is constant, the $I_c(H)$ dependence represents the well-known Fraunhofer pattern [2, 3]:

$$I_c(H) = I_c(0)\left|\frac{\sin(\pi\Phi/\Phi_0)}{\pi\Phi/\Phi_0}\right|, \tag{A1}$$

where $I_c(0)$ is the maximum supercurrent across the junction at $H = 0$.

However, in the case of a strong non-uniform supercurrent density, an alternative approach is needed. Let us consider it in more detail. Supercurrent density $J_s$ across the barrier sandwiched between two superconducting Nb electrodes with quantum phases $\theta_1$ and $\theta_2$ depends on their difference $\varphi = \theta_1 - \theta_2$. Alternatively, when a direct current of a certain magnitude passes through such a junction, the $\varphi$ value adapts to it according to the sinusoidal current-phase relationship between $J_s$ and $\varphi$: $J_s = J_c \sin\varphi$. If the magnetic field $\mathbf{H}(H,0,0)$ is applied along the $x$ direction parallel to the layers in the $xy$-plane (Fig. 1a in the main text), then the gauge-invariant phase difference $\varphi$ becomes dependent on $y$ (we use the Landau gauge), and the critical supercurrent $I_c$ through the junction can be found by maximizing the Josephson current $I_s$ [27]:

$$I_c(H) = \max|I_s(H)| = \left|\int_{-w/2}^{w/2} J_s(y)\exp(2\pi itHy/\Phi_0)dy\right|. \tag{A2}$$

Here $J_s(y)$ is the current density profile at point $y$ integrated over an in-plane perpendicular direction, $w$ is the width of the junction, $\Phi_0$ is the flux quantum, and $t$ is the effective magnetic thickness. The latter parameter includes the thickness $d_B$ of the barrier (a weak link), possible effect of its magnetization described by the magnetic permeability $\mu$, and the depths of magnetic field penetration into both niobium S electrodes, which are determined by the London penetration depth $\lambda_L$ and the thicknesses $d_S$ of the corresponding superconducting electrodes. The general formula for $t$ was given in the main text as Eq. (1) and is repeated here for consistency of discussion:

$$t = \mu d_B + \lambda_{Nb} \tanh(d_{S1} / 2\lambda_{Nb}) + \lambda_{Nb} \tanh(d_{S2} / 2\lambda_{Nb}). \quad (A3)$$

When both thicknesses $d_{S1}$, $d_{S2}$ >> $\lambda_L$, we get the conventional result: $t = \mu d_B + 2\lambda_L$ [60], but in the opposite limit $d_{S1}$, $d_{S2}$ << $\lambda_L$, $t = \mu d_B + (d_{S1} + d_{S2})/2$. In Eq. (A3), superconductors are considered to be identical but of different thicknesses, $\mu$ is averaged over the weak link and therefore, it strongly depends on the multilayer magnetic texture. The magnetic flux $\Phi$ through the weak link is a product of the magnetic field $H$ and the sensing area $s = tw$.

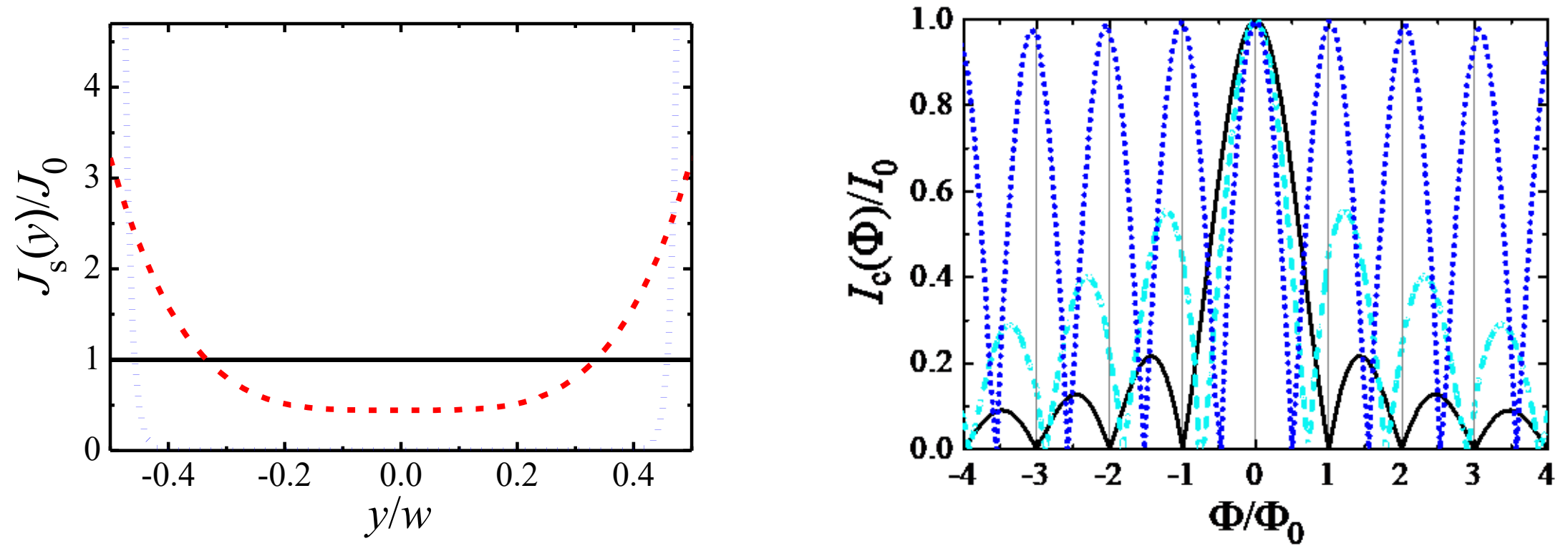

a b

**Fig. A2.** Effect of a non-uniform, symmetric $J_s(y) = J_s(-y)$ current density profile in a Josephson junction (a) on its $I_c(H)$ pattern (b), $J_0 = I_0/w$, $I_0$ is the maximum supercurrent through the heterostructure.

Using relation (A2), we will show what (at least qualitative) information about the current density profile in a Josephson junction can be obtained from the analysis of its $I_c(H)$ pattern shape. The general properties of the Fourier transform (A2) lead to the conclusion that *even* $J_s(y)$ spatial dependence necessarily leads to the nodes in the $I_c(H)$ curves [27]. In the case of a uniform distribution of the supercurrent $J_s(y)$ = const, we obtain the standard Fraunhofer characteristic (A1), black curves in Figs. A2(a) and (b). As the bulk contribution decreases and the edge charge flow increases, a smooth transition from the Fraunhofer curve to a periodic $I_c(H)$ pattern, reminiscent of the SQUID characteristics, takes place, see Fig. A2 for even in coordinate $J_s(y) = J_s(-y)$ spatial distribution of supercurrents through a weak link of the Josephson junction. It should be taken into account that the width of the central peak in the Fraunhofer curve is equal to $2\Phi_0$, while in SQUID-like dependences all periods are of the same value and equal to $\Phi_0$. This means that the nodes in the $I_c(H)$ patterns exactly correspond to integer flux quanta only in two extreme cases, completely uniform current distribution (A1) and that localized near the weak-link edges, *i.e.*, SQUID-like configuration, black and blue curves in Fig. A2. Otherwise, due to the presence of edge contributions that prevail over bulk ones, the formally defined period of nodes will be greater than that calculated by the formula $\Delta H = \Phi_0/s$, the red curve in Fig. A2.

Let us consider the extreme case of highly dominating edge currents having the magnitudes different for the two sides of the junction, i.e., $J_s(y) \neq J_s(-y)$. When the current distribution is symmetrical in coordinate, the nodes in the $I_c(H)$ pattern remain with a clear period determined by the magnitude of the magnetic field quantum $\Phi_0$; the black curve in Fig. A3, which is identical to the blue curve in Fig. A2. An increase in the asymmetry of the current density profile leads to an upward shift of the $I_c(H)$ curve along the axis of the maximum supercurrent $I_c$ with the *fixed* period $\Delta H = \Phi_0/s$, see Fig. A3. Note also a smoothly decaying envelope in the $I_c(H)$ dependences.

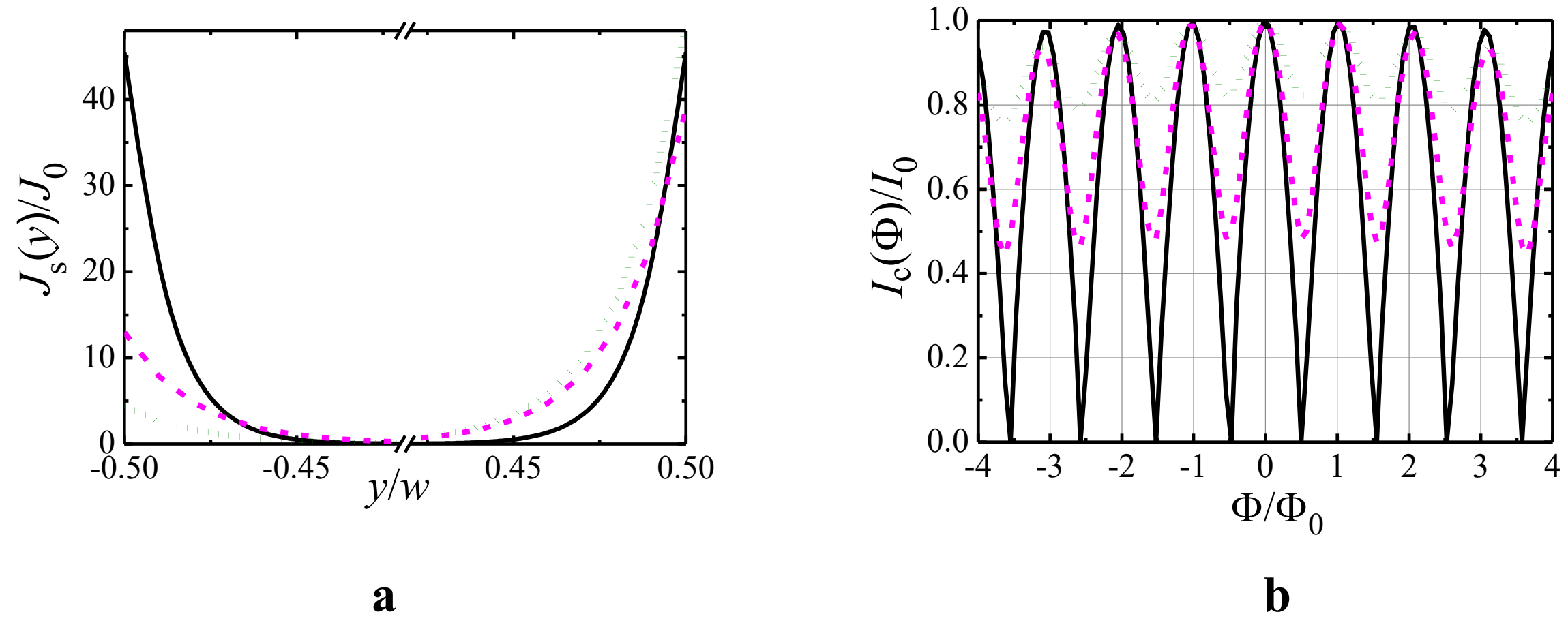

**Fig. A3.** Effect of a non-uniform, non-symmetric $J_s(y) \neq J_s(-y)$ current density profile in a Josephson junction (a) on its $I_c(H)$ pattern (b), $J_0 = I_0/w$, $I_0$ is the maximum supercurrent through the heterostructure.

Summarizing the results of numerical simulations shown in Figs. A2 and A3, we point out two main features of the Josephson spectroscopy characteristics $I_c(H)$, which qualitatively indicate specific spatial distribution of currents through the weak link in a Josephson device:

- The dominance of near-edge localized currents through a weak link in Josephson junctions leads to the transformation of the standard Fraunhofer $I_c(H)$ pattern into a periodic SQUID-like dependence.
- The presence of nodes in the $I_c(H)$ curves is a consequence of the symmetrical current density profile, while the appearance of a *current background* (a shift of the $I_c(H)$ characteristic upward along the maximum supercurrent $I_c$ axis) and the presence of a slowly decaying envelope indicate the asymmetry of the current distribution across the width of the sample.
- The period $\Delta\Phi$ of the $I_c(H)$ oscillations is exactly equal to $\Phi_0$ only when a non-chiral charge flow is localized inside extremely narrow regions adjacent to the edges of a Josephson junction.

Dynes and Fulton [27] considered Eq. (A2), relating the current density distribution and the magnetic interference pattern as a Fourier transform, and developed a mathematical procedure for determining the $J_s(y)$ function from experimental $I_c(H)$ data. Its more sophisticated version involves the Hilbert-transform technique that significantly complicates the calculation while a simpler approach [27] is based on the assumption that $J_s(y)$ is a nearly even function, which is not our case. In this work, we have adopted the following simplified strategy: we guess the approximate form of this dependence using the general principles of the Fourier transform [27], select a suitable mathematical equivalent with several unknown parameters, and find them with the least squares fitting method. This approach allows us to control the calculations at each stage and verify the adequacy of the reconstructed $J_s(y)$ dependence by matching the patterns obtained using Eq. (A2) with measured $I_c(H)$ curves.

## Appendix D. A short overview of our previous results [21, 61, 62]

### D1. SNINS junctions

We start with the analysis of conventional Josephson junctions obtained by the so-called niobium technology, in which an additional Al film deposited on the bottom Nb electrode is oxidized, creating about 1 nm thick tunnel barrier [63]. Since the Al overlayer is not oxidized completely, there is a remaining aluminum under the $AlO_x$ layer. To protect the upper Nb electrode from oxidation, we added another thin Al film on top of the resulting barrier. Note also that the additional Nb wiring layer increased the thickness of the top Nb electrode to about 500 nm. An example of the $I_c(H)$ dependence for a representative 10 μm × 10 μm Nb(117)Al/$AlO_x$(6.0)Al(3.1)/Nb(520) Josephson junction is shown in Fig. A4. The numbers in parentheses indicate nominal thicknesses of the respective layers in nm; as for Al/AlO$_x$, this is the thickness of the Al overlayer before thermal oxidation.

We fit the experimental $I_c(H)$ dependence with the expression (A1) using the maximum supercurrent value in the experimental curve $I_c(0) = 3.05$ mA (note that the maximum supercurrent is observed not exactly at $H = 0$ due to the self-field effect). The period $\Delta H = 15.4$ Oe of the Fraunhofer-like pattern for the junction width $w$=10 μm corresponds to the sensing area $s = \Phi_0/\Delta H = 1.34$ μm$^2$ and the effective magnetic thickness $t = s/w = 134$ nm. Next, we assume that the bottom S electrode includes about 5 nm-thick proximitized non-oxidized Al layer, while the top S electrode comprises 3.1 nm-thick Al layer deposited on top of about 1 nm-thick AlO$_x$ barrier. As a result, we obtain the following relation that defines $\lambda_{Nb}$ in an implicit form: $\lambda_{Nb}\tanh(d_{S1}/2\lambda_{Nb}) + \lambda_{Nb}\tanh(d_{S2}/2\lambda_{Nb}) = 133\ \mathrm{nm}$ with $d_{S1} = 122$ nm and $d_{S2} = 520$ nm, from which we deduce $\lambda_{Nb} = 82$ nm in agreement with previous works [57, 58]. This value of $\lambda_{Nb}$ will be used in the calculations below.

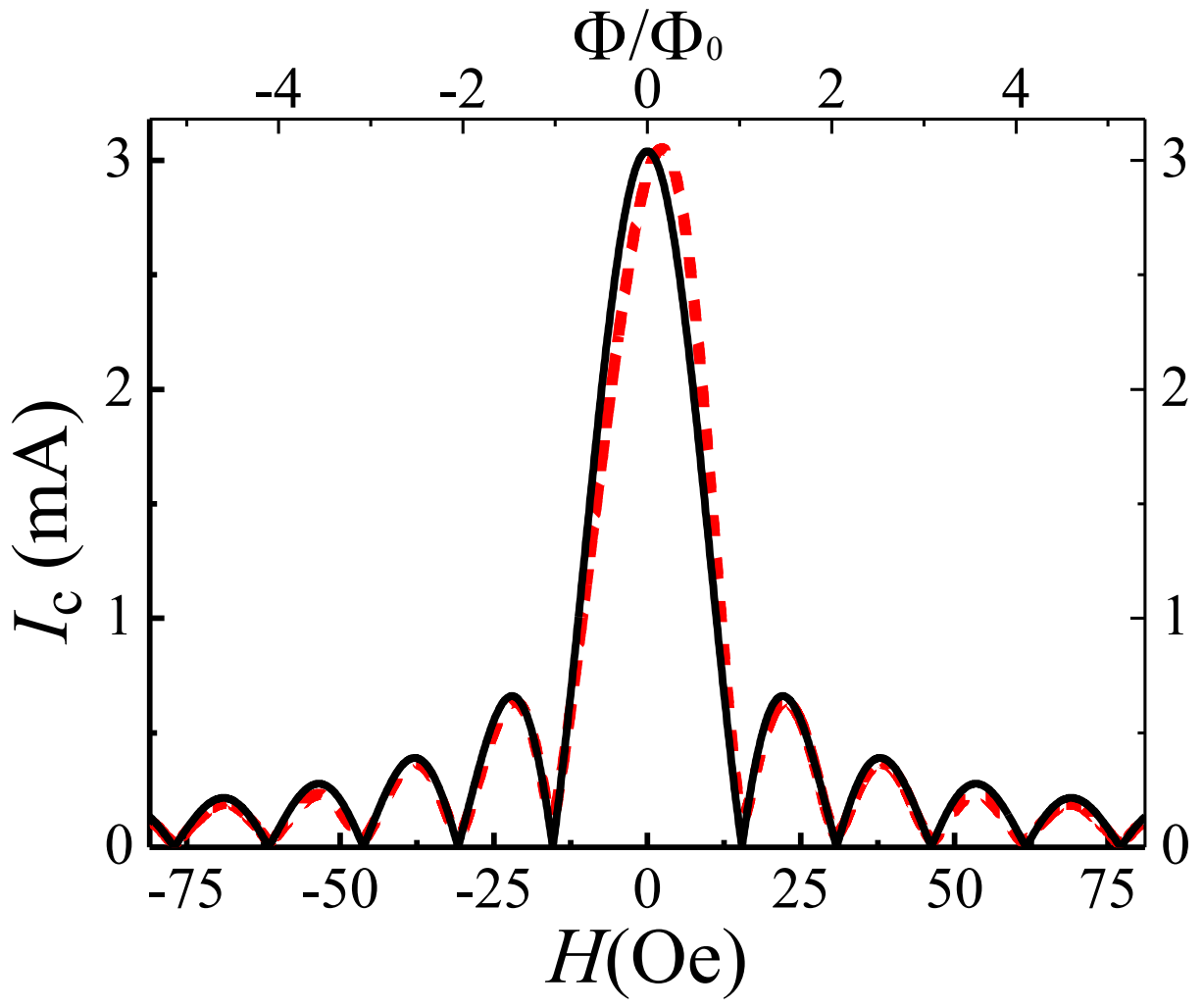

**Fig. A4.** $I_c(H)$ pattern (red dashed line) measured for a representative 10 μm × 10 μm Nb(117)Al/AlO$_x$(6.0)Al(3.1)/Nb(520) Josephson junction at $T = 4.2$ K. Black solid line shows the Fraunhofer $I_c(H)$ pattern (A1) for two fitting parameters: $I_c(0) = 3.05$ mA and the sensing area $s$ = 1.34 μm$^2$.

### D2. SN(F/N)$_n$I(N/F)$_m$NS junctions [61, 62]

Josephson junctions comprising ferromagnetic (F) metals have attracted a lot of interest due to the wide variety of new physical phenomena caused by the interaction of antagonistic electron orderings and great potential of such devices for practical applications [60]. Due to the exchange splitting in the F layer, the Fermi wave vectors are different in majority $k_F^{\uparrow}$ and minority $k_F^{\downarrow}$ spin bands and the Cooper-pair correlation function oscillates in space with the wave vector $Q = k_F^{\uparrow} - k_F^{\downarrow}$. It leads to damped oscillations in the dependence of $I_c$ on the F-layer thickness with π coupling regions corresponding to the phase difference of π for an SFS junction in the ground state. Most studies on S/F-based compounds have been performed on devices containing only a single F layer and having small enough lateral dimensions to be single-domain [57]. We have reported characteristics of SN(F/N)$_n$IN(F/N)$_m$S Josephson junctions with $n$, $m$ = 1 to 3 [61]. The devices with $n = m = 1$ and $n = 1$, $m = 2$ displayed distorted Fraunhofer patterns corresponding to the presence of a half of the flux quantum in the junction:

$$I_c(H) = I_c(0)\left|\frac{\sin \pi(\Phi/\Phi_0 \pm 1/2)}{\Phi/\Phi_0 \pm 1/2}\right| \tag{A4}$$

Due to the domain structure of the F interlayers, the half-flux shift was stochastic and switched between the two half-vortex states with emission of an integer vortex (Fig. A5).

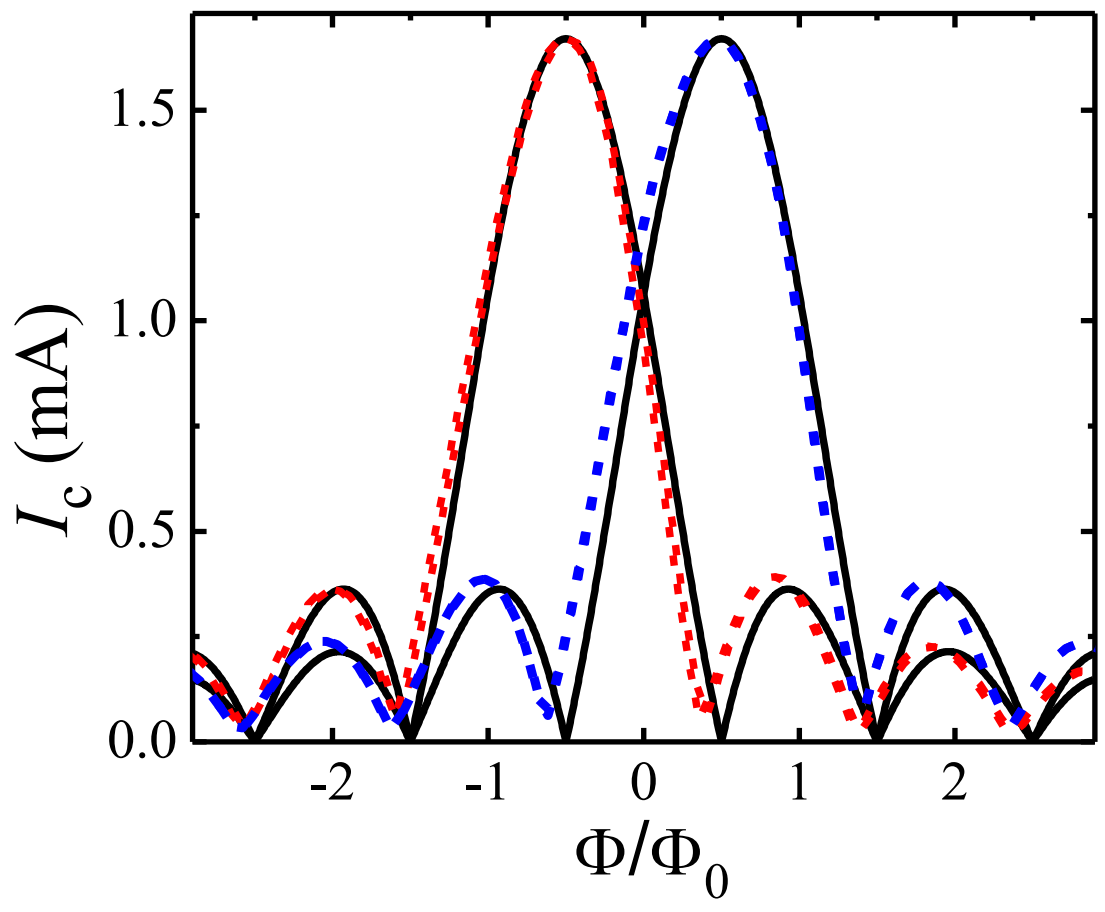

**Fig. A5.** $I_c(\Phi)$ patterns (red and blue dashed lines) measured for a representative 10 μm × 10 μm Nb(120)Al(3.1)Ni(1.3)Al/AlO$_x$(3.1)Al(3.1)Ni(1.3)Al(3.1)Nb(530) Josephson junction where the period of the magnetic field was about 11.1 Oe. The black solid lines show shifted by a half of a flux quantum to the left and right Fraunhofer patterns (A4), $T$ = 4.2 K [61].

The effective magnetic thickness $t$, which corresponds to the oscillation period $\Delta H$ = 11.1 Oe in Fig. A5, equals to $\Phi_0/(\Delta H\, w)$ = 186 nm. Let us calculate this value using the same procedure as above and including, as before, the Al layers adjacent to the Nb electrodes into the S thicknesses. For $\mu$ = 1, we get $t$ = 142.1 nm that is significantly less than the value of 186 nm obtained from the experiment. Since the agreement for the SNINS junctions was excellent, see above, the revealed disagreement could be associated with the presence of the magnetized Ni(1.3)Al/AlO$_x$(3.1)Al(3.1)Ni(1.3) region. Introducing for it $\mu \approx 5$, we obtain $t$ = 177.3 nm in good agreement with the experimental value of 186 nm. Of course, the latter estimate is rather rough, since the real distribution of internal magnetic fields is much more complex. However, the reached agreement qualitatively indicates the need to consider the magnetization of an F/N-based barrier when estimating the value of the effective magnetic thickness $t$. Of course, its contribution becomes significant only in the case of a large number of F/N bilayers, whereas for a single and ultrathin magnetic interlayer it can be ignored, as in the work [58].

In the SN(F/N)$_n$I(N/F)$_m$NS devices with $n = m = 3$, a typical $I_c(H)$ dependence consists of a background current with a large modulation period, and a Fraunhofer-like pattern with a small modulation period on top of the background current [61]. In Fig. A6, we show the experimental data for $n = m = 4$ where the zero-field peak totally disappears and the maximum supercurrent versus field dependence includes rapid oscillations with an envelope that is also a periodic function, but with a period several times larger [62].

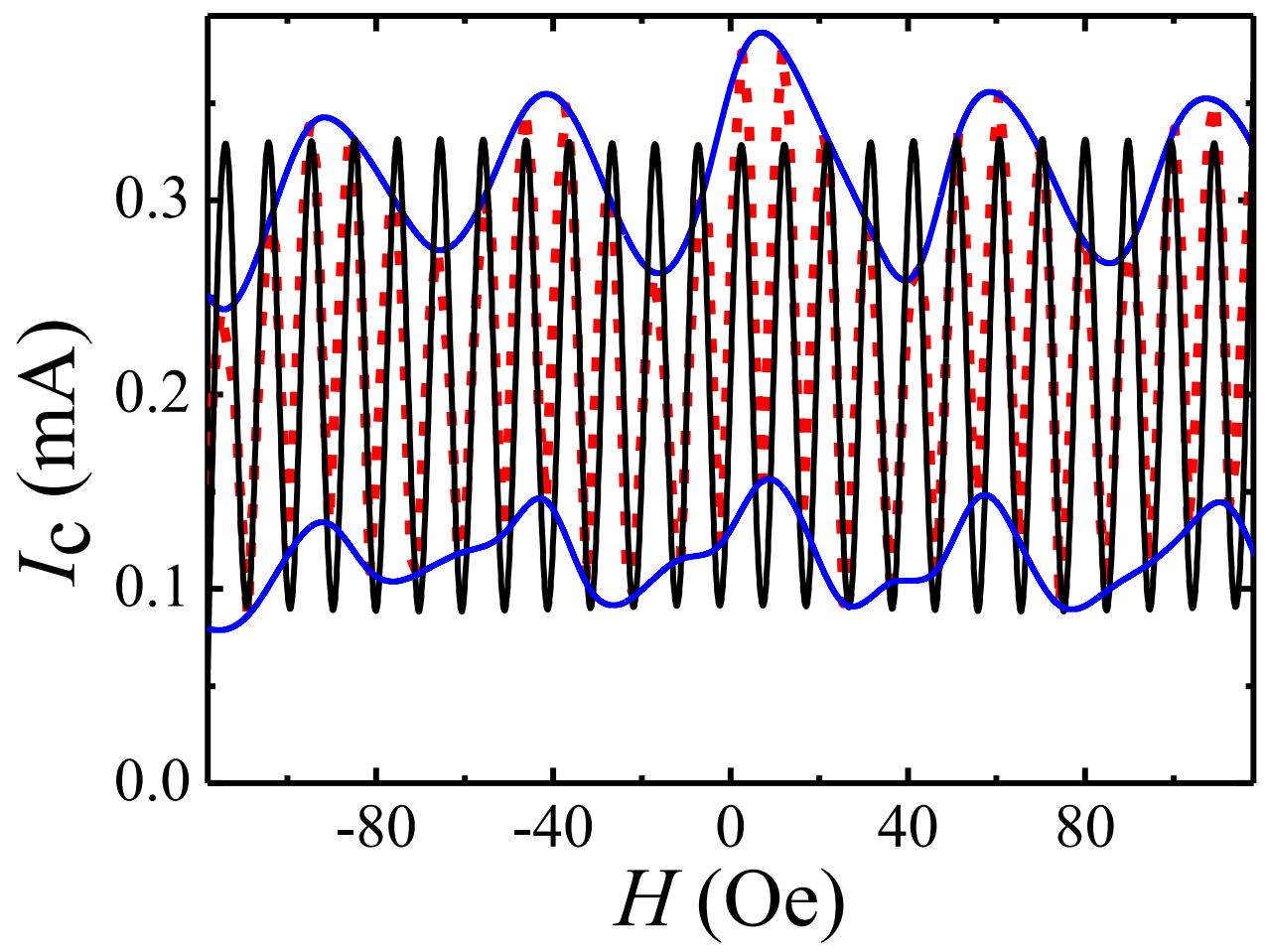

**Fig. A6.** $I_c(H)$ pattern (red dashed line) measured for a representative 10 μm × 10 μm Nb(120)[Al(3.1)Ni(1.2)]$_4$Al/AlO$_x$(3.1)[Al(3.1)Ni(1.2)]$_4$Al(3.1)Nb(70)/Nb(470) Josephson device at 4.2 K. The black solid line shows a periodic function $I_c$(mA) = 0.21+0.18·sin(2π$H$/Δ$H$) with the period Δ$H$ = 9.7 Oe. The blue lines are the envelopes of the measured $I_c(H)$ characteristic [62].

The measured $I_c(H)$ curve shown in Fig. A6 strongly deviates from the Fraunhofer dependence (A1). It is the sum of several periodic functions shifted upwards by a value of about 0.2 mA that is comparable to the amplitude of the main oscillatory contribution. The junctions with a lower number of F/N bilayers have not revealed such a shift, which appears only for a larger number $n$. Fig. A7(a) demonstrates an example of the data for $n > 4$ with three main trends arising in the modified $I_c(H)$ patterns, namely, an increased background current, reproducibility of the oscillation period, and the presence of a smooth envelope slowly decreasing as the field grows. Despite the difference in the osciilation amplitudes and the level of the background current, the oscillation period Δ$H$ = 13.0 ± 0.3 Oe is well reproducible in the $I_c(H)$ tracks for six nominally identical 5 μm × 5 μm Nb(120)[Al(3.1)Ni(1.2)]$_{10}$Al/AlO$_x$(3.1)[Al(3.1)Ni(1.2)]$_{10}$Al(3.1)Nb(470) junctions.

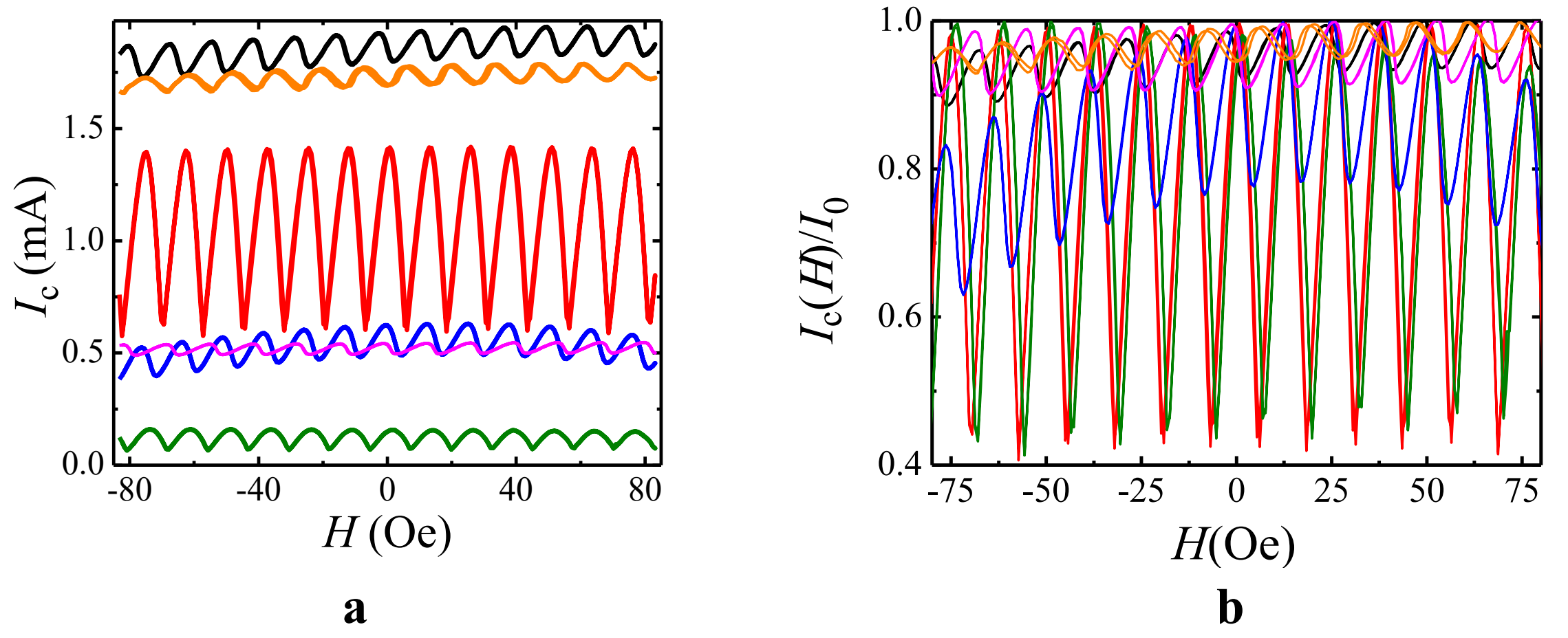

**Fig.** A7. $I_c(H)$ patterns for six 5 μm × 5 μm Nb(120)**[**Al(3.1)Ni(1.2)**]**$_{10}$Al/AlO$_x$(3.1) **[**Al(3.1)Ni(1.2)**]**$_{10}$Al(3.1)Nb(470) junctions measured at 4.2 K in a relatively smaal range of an in-plane applied magnetic field.

Using the conventional formula for the effective magnetic thickness $t = \Phi_0/(\Delta H\, w)$ and $\Delta H$ = 13.0 Oe, we get $t$ = 320 nm. Its estimates obtained by using the known thicknesses of the layers forming the heterostructure and $\lambda_{Nb}$ = 82 nm are $t$ = 220 nm for $\mu$ = 1 and $t$ = 565 nm for $\mu \approx 5$. If we normalize the measured characteristics to their maximum values, as in Fig. A3, then half of the samples demonstrate nearly complete asymmetry with a supercurrent that flows along one boundary. In other samples, the asymmetry is preserved, but not to the same extent. Following the conclusions of the work on disordered photonic alloys [32], the presence of robust edge modes with chiral properties can be assumed in our multilayers as well. However, in contrast to the publication [32], the magnetic state of the ultrathin Ni planes could not be controlled, which led to the emergence of current-carrying states of different chirality. Sometimes, as in the three samples of Fig. A7 and the multilayer structure shown in Fig. 1 of the main text, one particular direction of chirality dominates. In other cases, the oppositely directed chiral contributions to the supercurrent partially compensate each other. If we follow the assumption of chirality of the edge currents, then the relation for the period $\Delta H$ from $\Phi_0/s$ transforms to $2\Phi_0/s$ (see the main text), the value of $s$ found from the experimental curves in Fig. A7 is doubled and becomes equal to 640 nm, in quite satisfactory agreement with $t$ = 565 nm calculated above for $\mu \approx 5$.

### D.3. SN(F/N)$_{10}$S junctions [21]

Let us show examples of Josephson junctions without an intermediate insulating I layer and with a number of F/N bilayers equal to ten. Fig. A8 demonstrates representative $I_c(H)$ patterns for the devices with lateral dimensions of 1.5 μm × 1.5 μm (upper curves) and 0.9 μm × 0.9 μm (lower curves). The interference patterns are oscillating with a period increasing with decreasing lateral sizes of the junction and an envelope that slowly decreases as absolute field values grow. The periods for the two samples are $\Delta H$ = 81.5 Oe for the smaller device and $\Delta H$ = 36.2 Oe for the larger device. It appears that for very small lateral dimensions of the devices, scaling of the oscillation period with size deviates from that expected for the conventional SIS junctions. A noticeable hysteresis was observed for magnetic field sweeps in the range of relatively high fields, see Fig. A8. This is more pronounced in compounds with smaller lateral dimensions, which is consistent with the general trend of decreasing coercivity with particle size [64].

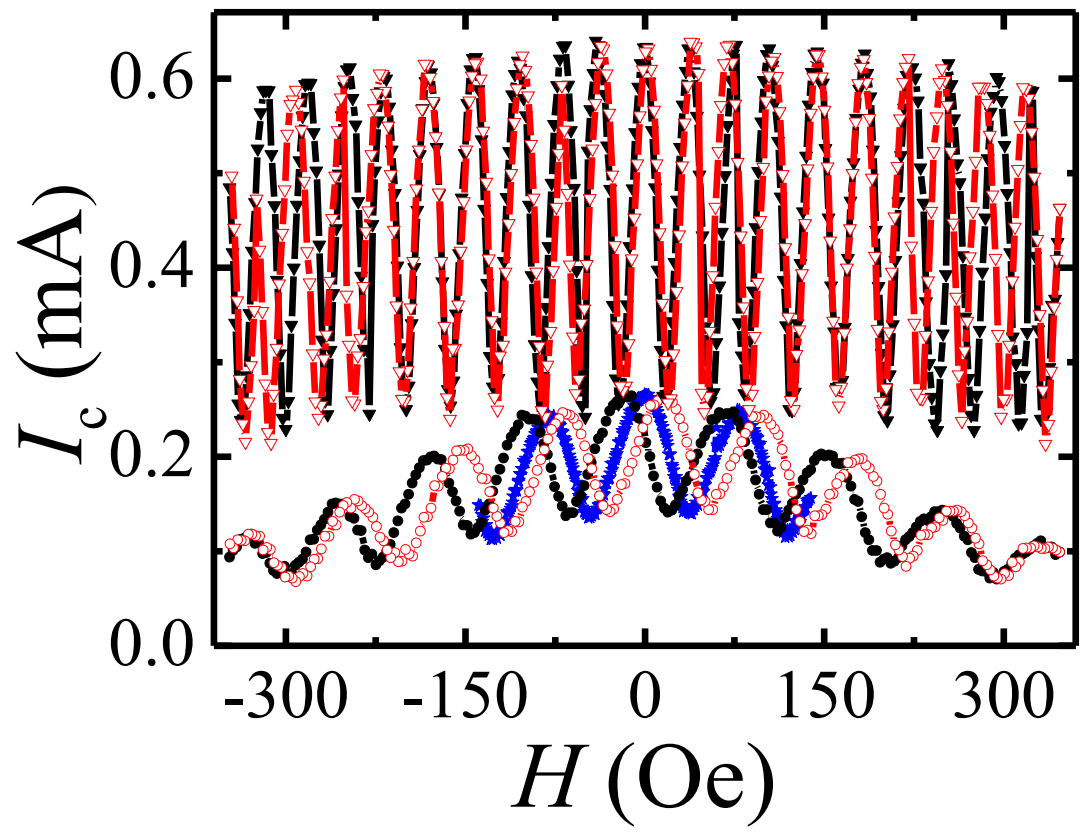

**Fig. A8.** Representative $I_c(H)$ SQUID-like oscillating characteristics for 1.5 μm × 1.5 μm (triangles, upper curves,) and 0.9 μm × 0.9 μm (circles and stars, bottom curves) Nb/(Al/Ni)$_{10}$Al/Nb junctions. Black and red symbols correspond to different directions of magnetic field sweeps within a relatively high-field interval. Non-hysteretic behavior (blue stars) is observed within a small range of field sweeping [21].

Generally, the SN(F/N)$_{10}$S junctions with the same layered structure and lateral sizes displayed approximately the same oscillation period, although considerable scattering of the oscillation amplitude and the level of the slowly changing with $H$ “background” supercurrent was observed. From the physics point of view, it appears that both types of multilayers, with and without an insulating interlayer, behave qualitatively identically. At the same time, due to a lower transparency of the junctions with the $AlO_x$ barrier compared to those without it, lateral dimensions of the former type of devices can be made larger, and therefore, they are easier to create and control. It was the main reason why we preferred to fabricate and study the samples with the $AlO_x$ barrier.